\documentclass{aa}  

\usepackage{graphicx}
\usepackage{txfonts}
\usepackage{lipsum}
\usepackage{orcidlink}
\usepackage{subcaption}         
\usepackage{lscape}             
\usepackage{placeins}           

\hypersetup{colorlinks=true,linkcolor=blue,citecolor=blue,urlcolor=blue }

\begin{document}

   \title{Inferring the delay time distribution of type Ia supernovae with \texttt{CAALAM}}

   \subtitle{I: SED modelling choices and their effect on the delay time distribution}

%
%
%

   \author{Aman N. Raju\orcidlink{0000-0001-9339-0789}\inst{1}\corrauth{aman.raju@nbi.ku.dk}        
        \and Jens Hjorth\orcidlink{0000-0002-4571-2306}\inst{1}
        \and Rados{\l}aw Wojtak\orcidlink{0000-0001-9666-3164}\inst{1}
        \and Christa Gall\orcidlink{0000-0002-8526-3963}\inst{1}
        }

   \institute{DARK, Niels Bohr Institute, University of Copenhagen, Jagtvej 155A, 2200 Copenhagen N, Denmark}

   \date{Received XX XX, 2026}

 
  \abstract
   {A delay time distribution (DTD) quantifies the lag between progenitor formation and the subsequent explosive event. The DTD for a sample of Type Ia supernovae (SNe Ia) from a survey can be modelled through relative rates and the star formation histories (SFH) of all individual galaxies. Modern wide-field surveys require computationally efficient methods to model the DTD without modelling large samples of non-host galaxies.}
   {We model the DTD for a volume limited sample from SNe Ia from the Zwicky Transient Facility. The aim is to study the effect of varying SED modelling assumptions, including stellar templates, dust attenuation, SFH models, and photometric coverage.} 
   {We present a new modelling framework, the Conditional Astral Ancestry Latency Analysis Machine (\texttt{CAALAM}), to infer the DTD of a sample of SNe Ia from the posterior of their host galaxy SFH.}
   {A model consisting of a standard stellar template, a non-parametric SFH and a flexible dust attenuation law, is favoured by the data and predicts a non-parametric DTD peaking at 100--330 Myr. When modelled as a power law, the DTD has a characteristic exponent of $-1.30 \pm 0.04$. The exponent is sensitive to dust attenuation, photometry, or SFH. When the DTD is convolved with the SFHs of the host galaxies, we find a bimodal progenitor age distribution, with $\sim60\%$ of progenitors with ages around 1.1--3.6 Gyr and the remaining progenitors older than 3.6 Gyr. We show a correspondence between the modes of the distributions of SN Ia light curve stretch and the progenitor ages.}
   {Our inferred DTD implies that most SNe Ia ($\sim 60\%$) occur promptly within 330 Myr of progenitor formation. However, the implied progenitor ages from the DTD convolved with the SFHs from our sample are significantly older. This suggests that the DTD might be a mixture of a prompt and an extended progenitor channel.}


   \maketitle
    \nolinenumbers

\section{Introduction}

Type Ia supernovae (SNe Ia) are important probes of cosmology. The relation between properties such as the peak brightness, rate of decline and colour of their light curves allow for standardization of their absolute magnitude \citep{Rust1974,Pskovskii1977,Phillips1993,Tripp1998}. This facilitates inference of cosmological distance scales, leading to discoveries such as the accelerated rate of expansion of our universe \citep{Riess1998, Perlmutter1999}. In the era of wide field surveys that monitor large areas of sky every night, such as the Dark Energy Survey Supernova Program \citep{DES_Supernova}, Zwicky Transient Facility \citep[ZTF,][]{ZTF_Overview}, the Vera C. Rubin Legacy Survey of Space and Time \citep[LSST,][]{LSST_Overview}, the All-Sky Automated Search for SuperNovae \citep{ASAS-SN_1,ASAS-SN_2}, Asteroid Terrestrial-impact Last Alert System \citep{ATLAS}, and Young Supernova Experiment  \citep{YSE} among others, an unprecedented number of SNe Ia are detected across wider and deeper areas than previous samples.

While it is known that SNe Ia are thermonuclear explosions of white dwarfs (WD) in binary systems, the precise nature of the companion star and mechanism for the explosion is still a subject of study \citep[e.g. see the reviews of ][and references therein]{Maoz_Review, Ruiter_2025_review}. Some of the leading progenitor models include the mass transfer from a non-degenerate companion (single degenerate scenarios, SD) and mergers of a double WD system (double degenerate scenarios, DD). These models can predict different timescales from progenitor formation to supernova, as well as impact the shape of the observed light curves and intrinsic luminosity, consequently affecting the accuracy of the standardization of their peak magnitudes \citep{SD_Model_SN,Greggio_DTD,Ruiter2009,Mennekens_DTD_Sim, DTD_Simulation_26}.

In modern cosmology with large samples of SNe Ia, errors in the standardization of their peak magnitude will be driven by systematic uncertainties rather than statistical uncertainties. Previous studies found that residuals in standardization possibly correlate with properties of the host galaxy such as stellar mass, metallicity, morphology, colour, and local environment \citep{Kelly_2010_Hosts,DLR_Paper_1, Ginolin_Color_Env,Ginolin_Stretch_Env}. This effect is taken into account in modern cosmological analyses in the form of environmental steps \citep{PantheonPlus,Ginolin_Stretch_Env,DES_Mass_Step}. The fundamental cause of this is still unknown. \cite{Briday_22_lsfr} found that the local specific star formation rate (sSFR) around the SNe Ia is a good tracer of SNe Ia peak magnitude, hinting at the progenitor age being a possible driver of this difference. Other studies found evidence for distinct channels characterizing the delay between the progenitor system formation and the SNe Ia, including the prompt channel where a SNe Ia occurs on timescales of 100s of Myrs after progenitor formation and a delayed channel where the SNe Ia can occur many Gyr after progenitor formation \citep{Manucci_2006_prompt_tardy,Childress_2014}. Thus, a detailed understanding of the delay time between progenitor formation and the SNe Ia and its subsequent effect on observed SNe Ia peak magnitude and light curves can reduce systematic uncertainties in the peak magnitude standardization.

The delay time distribution (DTD) of SNe Ia is a function that encodes the probability of a SN Ia occurring at a given delay time for a unit of stellar mass during a star formation episode \citep[][see Section \ref{subsec:DTD_modeling}]{Greggio_DTD,Maoz_2014_review}. When convolved with a star formation history (SFH) of a host galaxy, the DTD predicts the present day SNe Ia rates of the host galaxy. Previous studies have used stellar population synthesis codes to model DTDs for a range of progenitor scenarios \citep{Greggio_DTD,Ruiter2009,Mennekens_DTD_Sim,wang_sps_dtd,DTD_Simulation_26}. DD models predict that the DTD can be approximately described by a power law with a slope of $-1$, with fairly general underlying physics \citep[see][ for a detailed discussion]{Maoz_Review}. 

Using host galaxy star formation histories and observed SNe Ia rates has been an important method to infer a DTD from observations. Previous studies have compared observed SN Ia rates to the cosmic star formation history to infer a cosmic DTD \citep{Manucci_2006_prompt_tardy,Strolger_2020,Wiseman2021, Kim_26_Ages}. Other studies have used the SFHs of individual galaxies from spectral energy distribution (SED) modelling or from proxies for the SFH from galaxy colours, metallicity evolution models, or semi-analytical SFH models and determined a hierarchical universal DTD \citep{Bradt2010,Maoz_2011,Childress_2014,Color_DTD_Heringer,Strolger_2020,Nonparametric_DTD,Ca_rich_transient_DTD,Dwarf_Galaxy_DTD,Titan_SN_DTD}. Most studies have found DTDs consistent with a power law with a slope of around $-1$, favouring the predictions from DD scenarios. However, certain studies have found that the DTD might have a steeper slope. \cite{Sand_steep_DTD} and \cite{Color_DTD_Heringer} found DTDs that favoured a power law slope around $-1.5$ at large delay times. \cite{Dwarf_Galaxy_DTD} measured the DTD power law slope in dwarf galaxies to be closer to $-2$. There is also increasing evidence that the SN Ia rates may be dependent on other environmental factors such as the metallicity of the progenitor or the presence of the host galaxy within clusters \citep{LSS_excessive_sn,Dwarf_Galaxy_DTD,DTD_Simulation_26}.

The \cite{Maoz_2011} method infers the DTD based on a SNe Ia survey over a fixed volume. This is done by maximizing the likelihood of each galaxy hosting a number of SNe Ia (which can be 0, 1, or more in rare cases) over the course of the survey which is dependent on the SNe Ia rate of the galaxy from the convolution of the galaxy's SFH and the DTD. This requires modelling the SFH, both of galaxies that did and did not host SNe Ia during the given survey. With the increasing area covered by modern transient surveys, modelling the SFH of all galaxies within the area is challenging. 

The inference of galaxy SFH, along with other parameters derived from SED modelling has improved in the previous decades. Photometry being available from ultraviolet (UV) to near-infrared (NIR) wavelengths for a larger number of galaxies can better constrain contributions of young and old stellar populations and dust attenuation to the galaxy's SED \citep{Blast,Unions_SNeIa,Titan_SN_DTD,Frankenblast}. Modern SED modelling algorithms can determine a range of likely SFHs and related properties for a given galaxy in a Bayesian manner rather than being constrained by pre-defined templates. The \cite{Leja2019} continuity non-parametric SFH, along with two other non-parametric SFHs, can model a wider range of possible SFHs without imposing artificial constraints through an assumed functional form. In a similar vein, analogous non-parametric DTDs \citep{Maoz_2011,Maoz_2012,Nonparametric_DTD} also provide more information on the true shape of the DTD, including whether a power law approximation is justified.

In this series of papers, we present \texttt{CAALAM}\footnote{Conditional Astral Ancestry Latency Analysis Machine}, a modelling framework that infers parametric and non-parametric DTDs from host galaxy SFHs, through a novel conditional extension of the \cite{Maoz_2011} method. We apply \texttt{CAALAM} on the recently released ZTF SN Ia DR2 \citep{ZTF_SN_overview} sample (hereafter referred to as the ZTF sample). The ZTF sample is one of the largest samples to date of spectroscopically confirmed SNe Ia within a well-defined unbiased volumetric limit. The study of the delay times of these SNe Ia can unlock crucial insights about progenitors and constrain the DTD at low-redshifts where $K$-corrections do not dominate, with the potential to determine how it may evolve with redshift, and study the effect of the progenitor channel on observed SN Ia properties. In this paper, we describe our new method and study how assumptions built into SED modelling will affect modelled DTDs. In Paper II, we investigate the universality of the DTD with parametric and non-parametric formulations of the DTD, including different channels of SN Ia formation and the effects of host galaxy metallicity. In Paper III, we investigate how the channels correlate with observed SN Ia light curves and their standardization for cosmological analysis. In Paper IV, we demonstrate how to model SNe Ia beyond the volume limit and study the origin of subtypes of SNe Ia.

In section \ref{sec:Data}, we discuss our sample of host galaxies from the ZTF sample, our SED modelling methods and the conditional DTD modelling framework implemented in \texttt{CAALAM}. In section \ref{sec:Results}, we study the inferred DTD from \texttt{CAALAM}, the favoured baseline model over alternative SED modelling choices, and the resulting distribution of progenitor ages from the DTD. In section \ref{sec:Discussion}, we discuss these results, the validity of a universal DTD, and the implications for future studies. Finally, we summarize our conclusions in section \ref{sec:Conclusions}.

\section{Data and Methods}
\label{sec:Data}

In this section, we describe our sample of SNe Ia from the ZTF SNe Ia DR2 sample in subsection \ref{subsec:ZTF_SNe}, discuss SED modelling assumptions in subsection \ref{subsec:SED_model}, our DTD likelihood function in subsection \ref{subsec:DTD_modeling}, the inference of progenitor age from the DTD in subsection \ref{subsec:Age_infer} and finally the implementation of the modelling in subsection \ref{subsec:Modeling}.

\subsection{ZTF DR2 SNe Ia host data} \label{subsec:ZTF_SNe}

We choose the ZTF DR2 SNe Ia dataset \citep[ZTF sample]{ZTF_SN_overview} for the extensive number of spectroscopically classified supernovae, as well as the well-defined volumetric limit. Our sample cuts follow the recommendation from the ZTF sample \citep{Burgaz2025,Ginolin_Stretch_Env,ZTF_SN_overview} and are as follows.

We apply the volume cut of $z\leq0.06$ as per \cite{Amenouche2025}. Due to the low redshift limit, $K$-corrections are negligible and do not significantly affect the modelling of the host galaxy SEDs or the DTD. We choose only those SNe Ia that have well-sampled light curves and pass cosmological cuts by using the \texttt{lccoverage\_flag} provided in the associated catalogue. We also only use the SNe Ia that have associated host galaxies and redshifts. These host galaxies were determined by the directional light radius method \citep{DLR_Paper_1,DLR_Paper_2} as described in \cite{ZTF_SN_overview}. After these cuts, we have 997 SNe.

We use \texttt{HostPhot} \citep{HostPhot} to retrieve the photometry of the host galaxies through image co-adds from GALEX \citep{Galex_survey} FUV, NUV, PanSTARRS \citep{PanSTARRs_survey} $g$,$r$,$i$,$z$, and $y$, 2MASS \citep{2mass_survey} $J$, $H$, $Ks$, and WISE \citep{Wise_survey} $W1$ and $W2$, using the average PSF of each survey to adapt the aperture around the host galaxy. We use the deblended unWISE co-adds \citep{unwise_paper} for the WISE photometry. We require each host galaxy to have a detection in the PanSTARRS co-add to be considered in the analysis. This brings the sample down to 969 host galaxies.

Next, we correct for dust reddening due to the Milky Way. We utilize Galactic dust maps from \cite{MW_dustmap} for the Galactic extinction $E(B-V)$. Then we use the \texttt{dustmaps} \citep{dustmaps} implementation of \cite{Fitzpatrick_dust_modelling} to calculate $A_\lambda$ assuming a standard Milky Way $R_V$ of 3.1 \citep{Cardelli_Rv_MW}.

In order to reduce the Galactic dust correction bias dependent on the SED, we follow \cite{SED_dependent_MW_dust} and \cite{euclid_photoz} to construct a mock SED through cubic spline interpolation of the SED fluxes at the effective filter wavelengths. Then, we correct for Galactic extinction by the ratio of the flux as observed and the flux corrected for Galactic dust (from the $A_\lambda$ determination above). Then we apply this correction per band \citep[see][]{euclid_photoz}. We find that this method allows for better convergence during SED modelling and returns models that better match the observed photometry than only applying a \cite{Fitzpatrick_dust_modelling} correction.

\begin{figure}
    \centering
    \includegraphics[width=\linewidth]{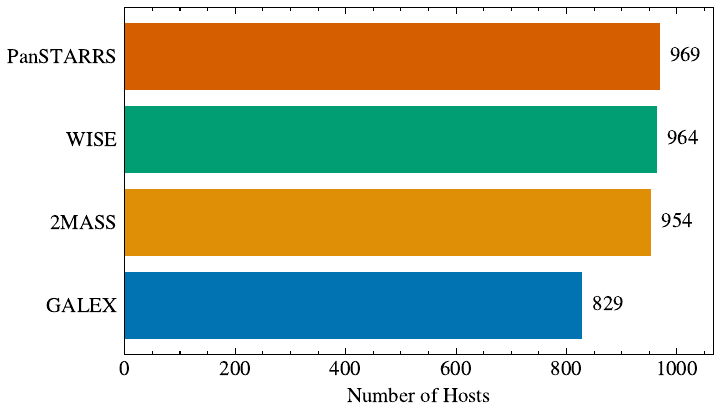}
    \caption{Number of host galaxies with photometry in each survey from those that have photometry detected from \texttt{HostPhot}, at least 4 filters, and detection in PanSTARRS.}
    \label{fig:survey_stats}
\end{figure}

Figure \ref{fig:survey_stats} shows that most of our sample of 969 hosts are present in PanSTARRS, 2MASS, and WISE. Around 100--200 galaxies do not have a detection in GALEX, which is either due to their quiescence or the shallow depth of GALEX. We do not expect this to cause a systematic bias as we utilize the full posterior SED models which account for the lack of GALEX data for these points in the form of sample uncertainties. 20 of the SNe Ia in this sample were classified by the ZTF sample release \citep{Dimitriadis_subtypes} as peculiar, while around 900 are specifically annotated as $\texttt{snia-cosmo}$ (i.e. standard normal or 91T type SNe Ia). We find that including or dropping the peculiar SNe Ia or only using the \texttt{snia-cosmo} SNe Ia do not affect the results in this paper, so we keep them in our sample.

\subsection{SED modelling}
\label{subsec:SED_model}

We use Bayesian Analysis of Galaxies for Physical Inference and parameter EStimation \citep[\texttt{BAGPIPES},][]{Bagpipes} to model the host galaxy SEDs. We choose many different SED assumptions to test their effect on the inferred DTD. In Table \ref{tab:Bagpipes_Priors}, we summarize the common choice of priors not associated with the SFH. The range of priors is intended to be as wide as possible to capture possible values similar to \cite{euclid_photoz} where applicable.

\begin{figure}
    \centering
    
    \begin{subfigure}{0.49\linewidth}
        \centering
        \includegraphics[width=\linewidth]{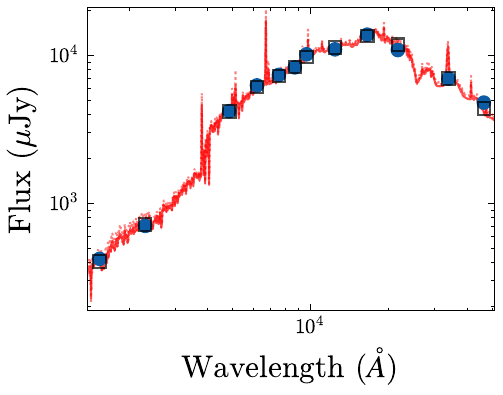}
        \label{fig:example_sed_fit_1}
    \end{subfigure}
    \begin{subfigure}{0.45\linewidth}
        \centering
        \includegraphics[width=\linewidth]{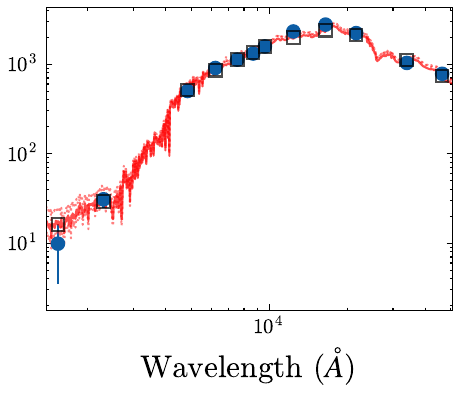}
        \label{fig:example_sed_fit_2}
    \end{subfigure}
    \begin{subfigure}{0.49\linewidth}
        \centering
        \includegraphics[width=\linewidth]{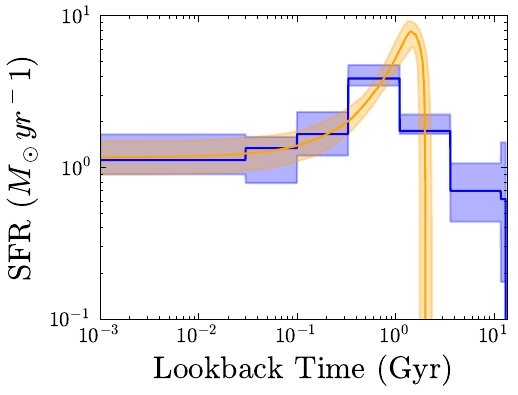}
        \label{fig:example_sfh_1}
    \end{subfigure}
    \begin{subfigure}{0.45\linewidth}
        \centering
        \includegraphics[width=\linewidth]{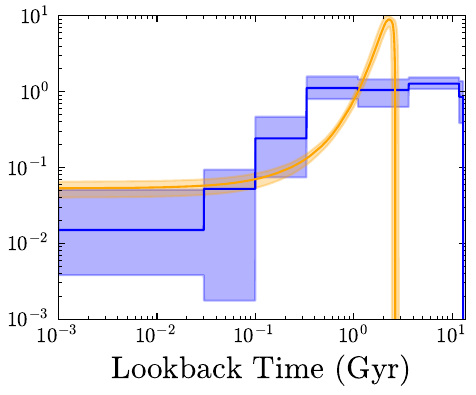}
        \label{fig:example_sfh_2}
    \end{subfigure}

    \caption{Example SED fits (top) and SFHs (bottom) for two example galaxies, one with a relatively higher SFR (left) and another with a lower SFR (right). The SED fits show the observed photometry (blue dots with error bars), 
    SED modelled spectra (red with 16th--84th percentile errors in dashed red lines), and the corresponding photometry (boxes with black outlines). The photometry is modelled from GALEX UV 
    bands to WISE MIR bands. The corresponding median continuity (blue) and delayed-$\tau$ (orange) SFHs associated with each galaxy are shown with their 16--84th percentile error bars.}
    \label{fig:sed_sfh_comparison}
\end{figure}

\texttt{BAGPIPES} creates modelled spectra, convolves them with filter curves to get the modelled photometry, and identifies which parameters maximize the Bayesian likelihood that the modelled spectrum matches the observed photometry through the \texttt{NAUTILUS} nested sampler \citep{nautilus}. Figure \ref{fig:sed_sfh_comparison} (top) shows an example of the modelled SED for galaxies with different star formation rates (SFRs).

We model the redshift of the host galaxy similar to \cite{Radek_ZTF_two_pop}. First, we correct the provided ZTF sample redshifts for the motion of the solar system through the Cosmic Microwave Background ($z_\mathrm{CMB}$) \citep{planck_results_1,planck_results_3}. However, the lowest redshift objects ($z\lesssim0.02$) may contain peculiar velocity offsets on the order of 300 km$\mathrm{s^{-1}}$ that could bias the host galaxy SED \citep{ZTF_Peculiar_Velocity,ZTF_SN_overview,Radek_ZTF_two_pop}. We correct $z_\mathrm{CMB}$ for these peculiar velocities with a cosmic flow model \citep{Cosmic_Flow}. We propagate an additional error of 300 km$\mathrm{s^{-1}}$. Finally, we use a Gaussian prior centred at the corrected redshift ($z_\mathrm{cosmo}$) and a standard deviation derived from the corrected error. We do not find any differences between the inferred DTD using $z_\mathrm{cosmo}$ and $z_\mathrm{CMB}$ but find a significant increase in Bayesian evidence ($\Delta \log Z > 10$) for DTD models using $z_\mathrm{cosmo}$, and thus adopt $z_\mathrm{cosmo}$ for the rest of our series.

We detail the different components included in our SED modelling including the stellar population templates, dust attenuation laws, and the SFH below.

\begin{table}
    \centering
        \caption{Priors shared among different SFH assumptions in our \texttt{BAGPIPES} SED modelling. All priors are logarithmic except for $\delta$ and $Q_{PAH}$}
    \begin{tabular}{|c|c|c|}
        \hline
        \textbf{Parameter} & \textbf{Prior Range} \\
        \hline
         Stellar mass ($M_\odot$) & $1$--$10^{13}$ \\
         \hline
         Metallicity ($Z_\odot$) & $0.01$--$2.5$ \\
         \hline 
         Dust attenuation, $A_V$ (mag) & $0$--$4$ \\
         \hline
         Attenuation curve steepening, $\delta$ & $-1.6$--$0.4$ \\
         \hline
         PAH fraction, $Q_{PAH}$ & $1$--$4.5$ \\
         \hline
         Starlight intensity limit, $U_{min}$ & $0.1$--$25$ \\
         \hline
         Fraction of stars at $U_{min}$, $\gamma$ & $10^{-4}$--$10^{-2}$ \\
         \hline
         Ionization parameter U & $10^{-4}$--$10^{-2}$ \\
         \hline
    \end{tabular}
    \label{tab:Bagpipes_Priors}
\end{table}

\subsubsection{Stellar population templates}

The choice of the stellar population templates is an important element in the result of the SED modelling. For our standard template, we choose the default \texttt{BAGPIPES} stellar templates which are 2016 versions of the \cite{BC03} stellar templates (hereafter referred to as BC03) with a Kroupa \citep{Kroupa_IMF} initial mass function (IMF). As an alternative, we also use Binary Population and Stellar Spectrum \citep[\texttt{BPASS}, ][]{BPASS_1,BPASS_2} v2.2.1 with the default IMF known as the \texttt{imf 135-300}. BPASS stellar population templates explicitly models binary stellar populations to match present day fractions.

The Kroupa and BPASS IMFs are similar with the main difference stemming from the IMF cut-off mass being 300 $M_\odot$ for BPASS, but 100 $M_\odot$ for BC03. The slope for stars more massive than 0.5 $M_\odot$ also differs slightly being $-1.3$ for Kroupa and $-1.35$ for the BPASS IMF. Other differences stem from the choice of metallicity and age grids provided with each of the templates. BC03 grids are finer than BASS grids for old stellar populations.

\subsubsection{Dust attenuation laws}

We adopt the \cite{Salim_Dust} law (hereafter referred to as the Salim attenuation law) for the line of sight dust attenuation within \texttt{BAGPIPES} as our standard dust attenuation law. This is a modification of the \cite{Calzetti_dust} attenuation law (hereafter referred to as the Calzetti attenuation law). The Salim attenuation law also allows for a modified UV bump strength, however we set this to 0 as in \cite{Meldorf2024}. The deviation of the attenuation slope from that of the Calzetti attenuation law is modelled with a power law with a characteristic parameter $\delta$. The choice of the prior on $\delta$ ($-1.6$ to $0.4$) comes from \cite{Meldorf2024}, which is slightly wider than the \cite{Salim_Dust} $\delta$ range of $-1.2$ to 0.4. In addition, we also model the dust attenuation through the Calzetti attenuation law as implemented in $\texttt{BAGPIPES}$ to test the effect of this assumption on the inferred DTD. This is equivalent to the Salim attenuation law with $\delta = 0$.

\subsubsection{Star Formation Histories}

We use the \cite{Leja2019} `non-parametric' continuity SFH as our baseline SFH (henceforth referred to as the continuity SFH). This parametrizes the SFH as a series of bins that are log-spaced in lookback time with each bin having a characteristic SFR. The adjacent bins have a SFR that is continuous through a Student-t prior with degrees of freedom $\nu = 2$ and the scale $\sigma = 0.3$. We adopt the same bins as \cite{Leja2019} with the minor exception of the final bin, where we assume that cosmic star formation begins at $z = 20$, thus decreasing the cut-off of the final bin by 200 Myr:

\begin{align}
\begin{split}
\label{eqn:bins}
    0 <\:  &t < 30\:  \mathrm{Myr} \\
    30 <\:  &t < 100\:  \mathrm{Myr} \\ 
    100 <\:  &t < 330\:  \mathrm{Myr} \\ 
    330 < \: &t < 1100\:  \mathrm{Myr} \\
    1100 <\:  &t < 3600\:  \mathrm{Myr} \\ 
    3600 <\:  &t < 11700\:  \mathrm{Myr} \\ 
    11700 <\:  &t < 13500\:  \mathrm{Myr}
\end{split}
\end{align}

We compare the continuity SFH to the delayed-$\tau$ model \citep[e.g.][]{delayed_tau_2,delayed_tau_1,Parametric_SFH} as an example of a parametric SFH model. This assumes that the SFH is described as ($t-t_0)/\tau$ exp(-$t-t_0$/$\tau$) at $t>t_0$, where $t$ is the lookback time, $t_0$ is the age, and $\tau$ is a characteristic time scale. We sample the age as $0.01$--$13.74$ Gyr and $\tau$ as $0.01$--$31.6$ Gyr, both on a log scale. Figure \ref{fig:sed_sfh_comparison} shows the different SFH that we have considered in our study. The delayed-$\tau$ is smoother than the continuity SFH. However, the continuity SFH can model different periods of star formation more flexibly.

We also note that we have experimented with the \texttt{dense\_basis} SFH \citep{Dense_Basis,dense_basis_package} in \texttt{BAGPIPES} that uses Gaussian processes and adaptive binning by mass percentile instead of lookback time. However, our subsequent DTD will be inferred over a series of common delay time bins across all of the host galaxies in our sample. Thus, we do not find \texttt{dense\_basis} suitable for this method. This is in contrast to \cite{Titan_SN_DTD} who used $\texttt{dense\_basis}$ with an assumed smooth power law DTD to model progenitor ages. However, \cite{Titan_SN_DTD} found that the continuity and \texttt{dense\_basis} SFH result in similar SED model posteriors. Thus, both methods yield similar results as shown in Section \ref{subsec:ages}.

\subsection{The Delay Time Distribution} \label{subsec:DTD_modeling}

Our aim is to infer and study the shape of the DTD, as the absolute normalization requires information about the absolute SN Ia rates which might not be available. Thus, we decompose the absolute DTD, $\psi$, into two parts, $\psi = \eta\phi$, where $\eta$ is the amplitude of the DTD derived from $\int \psi dt$ and $\phi$ is the normalized DTD, which can be interpreted as a probability density function as $\phi = \psi$ when $\eta=1$. For simplicity, we will refer to $\phi$ as the DTD, $\eta$ as the DTD amplitude, and $\psi$ as the absolute DTD. 

In the scope of this paper, we make the assumption that the absolute DTD is universal in both shape and amplitude, as in \cite{Maoz_2011}. We follow the \cite{Maoz_2011} method of inferring this universal DTD from the SFH of galaxies observed in a survey. While this can be broadly applied to any extragalactic transient phenomenon, we describe the process for a SN Ia survey. \cite{Maoz_2011} derives a likelihood function for a given DTD model by convolving it with a given SFH (which is a simple dot product if both are binned) to give an intrinsic SN Ia rate of a galaxy. Then, the intrinsic SN Ia rate is corrected for the observation time and detection efficiency of SNe Ia from each host galaxy in the survey (accounting for biases in cadence, luminosity, etc.) to give the expected number of SN Ia observed in each galaxy over the course of the survey. However, since the likelihood of detecting a SN Ia over the duration of a typical survey is very small, the expected number of SNe Ia is modelled as a Poisson process. The Poisson likelihood of the galaxy hosting a SN Ia is maximized through Monte-Carlo Markov Chain (MCMC) sampling, or nested sampling, to provide the posterior greatest likelihood DTD models.

The expected number of supernovae for the $i$th galaxy in a given survey is 
\begin{equation}
    \lambda_i = \frac{r_i\Delta t_i \epsilon_i}{1+z},
\end{equation}
where $\Delta t$ is the duration of time over which the galaxy is observed (assumed to be roughly similar for each galaxy throughout the ZTF survey), $\epsilon$ is the efficiency of detecting a SNe Ia from the galaxy (accounting for observational factors such as redshift, Galactic extinction, etc.), the factor of $(1+z)^{-1}$ accounts for the time dilation from the host frame to the observer frame \citep{Maoz_2012} and $r_i$, the intrinsic supernova rate, comes from 
\begin{equation}
    r_i = \sum_{j} m_{ij}  \psi_j ,
    \label{eqn:dtd_convolution}
\end{equation}
where $\psi_j$ is the value of the absolute delay time distribution in the $j$th bin and $m_{ij}$ is the mass formed by the $i$th galaxy in its $j$th bin. As the rate derives from a convolution, the bins for the DTD and the SFH are the same, but the DTD is in delay time and the SFH is in lookback time.

We also assume that every galaxy either has one SN Ia or none. There is one galaxy in our sample that has hosted two SNe Ia that we only include once in our model (the host of ZTF20abydrkl and ZTF20acpmgdz). Thus, the Poisson log-likelihood derived from the expectation value $\lambda_i$ is simply:

\begin{equation}
\begin{cases}
    \log P_\text{host}(n_i|\lambda_i) = \log \left(\lambda_i \right) - \lambda_i \\
    \log P_\text{non-host}(n_i|\lambda_i) = - \lambda_i
\end{cases}
\label{eqn:ind_likelihoods}
\end{equation}
and as each galaxy likelihood is an independent variable, the total log-likelihood is a sum over all the individual galaxy log-likelihoods
\begin{equation}
    \log P_\text{total} = \sum_{k=0}^K \log \left( \lambda_k\right) -\sum^N_{i=0} \lambda_i,
    \label{eqn:full_likelihood}
\end{equation}
where $k$ in an index which denotes galaxies that host a SNe Ia. Thus, the first term only comes from $K$ host galaxies, while the second term comes from all $N$ galaxies.

In \cite{Maoz_2011,Maoz_2012}, this method is performed in a given field, where the SFH of all galaxies is known, both those which happened to host a SN Ia during a survey and those which did not. However, it is computationally expensive, especially for a wide field survey, to model the SFH of all galaxies that did not host a SN Ia during the survey.

Here, we introduce our conditional DTD. Instead of maximizing the likelihood across all galaxies including those that did not host SNe Ia during a survey, we model the DTD that maximizes the likelihood that each of the $K$ galaxies in our sample hosts a SN Ia that is observed during the survey. This is the likelihood that all of our galaxies are observed as hosts, given that we have $K$ SN Ia in our sample and an assumed DTD, which is expressed through Bayes' theorem as:

\begin{equation}
    P(\text{are hosts}|K,\psi) = \frac{P(\text{are hosts},K|\psi)}{P(K)}.
    \label{eqn:bayes}
\end{equation}

\noindent The numerator is the joint probability that all of the host galaxies are indeed hosts. This is simply the sum of the log-likelihoods from Equation \ref{eqn:full_likelihood}. The denominator is the probability of seeing $K$ SNe Ia across the survey. The log-likelihood for the entire survey is the sum of all the independent likelihoods. As the sum of independent Poisson likelihoods is itself a Poisson likelihood with the expectation value $\Lambda = \sum^N_{i=0} \lambda_i$, the denominator is:
\begin{equation}
    \log P(K) = K\log \Lambda - \Lambda - \log K!
\end{equation} 

\noindent Substituting these likelihoods into Equation \ref{eqn:bayes} and taking the logarithm gives
\begin{equation}
\log P(\text{are hosts}\mid K,\psi) = \left(\sum_{k=1}^K  \log \lambda_k\right) - \Lambda - K\log \Lambda + \Lambda + \log K!
,
\end{equation}

\noindent where the sum is over the host galaxies only, while $\Lambda$ is derived from all galaxies. The $\Lambda$ term cancels out and we drop the $\log K!$ term as it is a constant and cannot be optimized by sampling. Finally, the log-likelihood is

\begin{equation}
    \log P(\text{are hosts}|K,\psi) = \sum_{k = 1}^K \log \left( \frac{\lambda_k}{\Lambda} \right),
\end{equation}

\noindent which is a multinomial likelihood over the observed hosts conditioned on the number of supernovae.

At this point, the model is equivalent to the Poisson likelihood model from \cite{Maoz_2011}. However, it cannot recover a unique $\psi$ as the amplitude $\eta$ cancels from both the numerator and denominator. Thus, we only model $\phi$. When comparing to \cite{Maoz_2012}, we take $\eta$ from their study. While other works have calculated $\eta$, \citep[e.g.][]{Maoz_2011,Maoz_2012,Strolger_2020, Wiseman2021}, the true value of $\eta$ is beyond the scope of this paper.

We fix $\Lambda$ to be the total SN Ia rate across only the host galaxies during the survey,

\begin{equation}
\Lambda
\sim
\Lambda_{\mathrm{host}}
=
\sum_{k=1}^{K}\lambda_k,
\end{equation}

\noindent which yields the final conditional log-likelihood

\begin{equation}
\log P(\text{are hosts}\mid K,\psi)
=
\sum_{k=1}^{K}
\log\left(
\frac{\lambda_k}
{\sum_{k'=1}^{K}\lambda_k'}
\right),
\end{equation}

\noindent where we now have a likelihood that maximizes the probability of the galaxies hosting SNe Ia during a given survey under an assumed DTD.

Under the assumption that our sample does not include any observational selection biases ($\epsilon = 1$), the conditional DTD is identical to the DTD as described in \cite{Maoz_2011} upto a normalization. This assumption is justified for our sample as it does not suffer from selection-based effects in the volume limit under the light curve and cosmological filters in Section \ref{subsec:ZTF_SNe} \citep{Amenouche2025}. In Appendix \ref{app:conditional_dtd_test}, we test the response of this framework on mock data and show that the modelled DTD is prior-driven and does not converge on random data, and it infers a flat DTD on realistic SFHs with no strong underlying DTD.

However, it should be noted that our likelihood depends on the sample of SNe Ia being considered. For example, the method will recover different DTDs for high and low stretch SNe Ia (as in \cite{Maoz_2012} and \cite{Wiseman2021}). Thus, a general DTD can only be derived from a sample with no selection biases, such as our sample from the ZTF sample volume limited sample with the appropriate cuts.

\subsection{Progenitor age inference from the DTD}
\label{subsec:Age_infer}

Equation \ref{eqn:dtd_convolution} shows that the present day SNe Ia rates in each host galaxy result from a sum over the product of the DTD and the stellar mass formed in each lookback time bin. Thus, each lookback time bin produces a fraction of the current SNe Ia rate. As $\phi$ is normalized to 1 and we sample a SNe Ia rate for each host galaxy, we define the probability of the $j$th lookback time bin producing the SNe Ia observed in a galaxy during a given SNe Ia survey as:
\begin{equation}
    p_{ij} =\frac{m_{ij}\phi_j}{\sum^N_{j=0} m_{ij}\phi_j}.   
\end{equation}

\noindent Then, we assign each lookback time bin a characteristic time defined by the start and end lookback times in each bin as $t_\mathrm{j} =\sqrt{t_\mathrm{start,j}t_\mathrm{end,j}}$. The expectation value of the characteristic time given $p_{ij}$ gives the progenitor age as:
\begin{equation}
    E[\mathrm{age}]_i = p_\mathrm{ij}\times t_j,
\end{equation}
\noindent which matches the method for calculating the progenitor ages from the SFH of a host galaxy and DTD from previous studies \citep{Childress_2014,wiseman_dtd_ages,chung_wiseman_response,Titan_SN_DTD}. The expectation value should be interpreted in the context of the delay time bins and under the knowledge that the SED modelling is sensitive to logarithmic variations in stellar population ages \citep{Ocvirk_stellar_log}. 

\noindent 

\subsection{Modeling}
\label{subsec:Modeling}

Our likelihoods for all the models in \texttt{CAALAM} are written in \texttt{numpyro} \citep{numpyro_package}, a \texttt{NumPy} \citep{numpy_paper} backend for \texttt{Pyro} \citep{pyro_package} powered by \texttt{JAX} \citep{jax2018github}. While we also include MCMC sampling through a No U-Turn Sampler \citep[][NUTS]{NUTS_Paper} sampler based on Hamiltonian MCMC in \texttt{CAALAM}, we choose to use nested sampling in our study. \texttt{jaxns} \citep{jaxns} is a nested sampling algorithm built in \texttt{JAX} and integrated into \texttt{numpyro}. This allows for models developed in \texttt{numpyro} to be reused in nested sampling. 

Using nested sampling allows us to calculate the evidence $Z$ of each model, which is useful for model comparison. Nested sampling is also able to estimate multi-modal posteriors as it samples posterior volume using live points drawn from the prior, guarding it against situations where the model is split between a multimodal posterior that can affect other MCMC models like NUTS \citep{Nested_Sampling}. We find that this does not affect the non-parametric DTD but will be important for the power law model and the functional models that we will adopt in future papers.

We infer the DTD with two assumptions in this study, through a non-parametric Dirichlet distribution or a power law. For the non-parametric method, we first sample the integrated value of the DTD $\phi_j\Delta t_j$ in each delay time bin from a prior Dirichlet distribution with $\alpha = 1$. Then, we divide each Dirichlet value by the width of each bin $\Delta t_j$ in order to infer the DTD value in each delay time bin $\phi_j$. We refer to this as the non-parametric DTD (in convention with similar SFH methods; e.g see \cite{Leja2019}). We choose the delay time bins similar to Equation \ref{eqn:bins} but drop the first bin of 0--30 Myr to set a minimum time for white dwarfs to form ($t_\text{min} = 30$ Myr). This is slightly lower than but in line with other studies such as \cite{Maoz_2012} and \cite{Wiseman2021}, where the minimum time is set to 40 Myr. Thus, we have 6 DTD delay time bins from 30 Myr to 13.5 Gyr, which implies 5 sampled parameters for the Dirichlet distribution.

Along with the non-parametric DTD, we also sample a binned power law DTD. This is formulated as follows:
\begin{equation}
    \phi_i = \int_{t_{i-1}}^{t_i} \left(t-t_{
    \text{min}} \right)^{-\alpha} dt,
\end{equation}
\noindent where the value of the DTD is determined by the integral of the power law from the start and end times of the characteristic bin and $\alpha$ is the characteristic power law index.

We sample $\alpha$ from a Gaussian prior of $\mathcal{N}(-1,2) $. For the $t_\text{min}$ parameter, we sample the value of $t_\text{min}$ between the lower bound of 30 Myr and upper bound of 13.5 Gyr, with a sigmoid function instead of a sharp transition at $t_\text{min}$.

An important hyperparameter for the nested sampler is the initial number of live points. We roughly follow the recommended number of $50\times(D+1)\times(N_\text{posterior})$ where $D$ is the dimensionality of the posterior and $N_\text{posterior}$ is the expected number of modes in the posterior \citep{jaxns}. Thus, we use 1000 live points for the non-parametric DTD model (5 parameters) and 500 live points (2 parameters) for the power law DTD model. Once the sampling has finished, we draw 1500 samples from the weighted posterior. We use \texttt{arviz} \citep{arviz} for posterior analysis. 

In order to measure the true uncertainty from both SED modelling and our sampling, we marginalize the likelihood over the full posterior of the \texttt{BAGPIPES} modelled SFH (500 posterior samples per host galaxy) in the sampling of the DTD thus propagating SED model uncertainties into the derived DTD. We use the derived DTD to infer progenitor ages post-hoc by drawing 10,000 random samples from both the SFH samples and the derived DTD and calculating the median progenitor age $E[\mathrm{age}]$ for each host galaxy. We report the median value for all quantities in this studies, with error bars derived from the 16 and 84 percentile posterior  values.

\section{Results}
\label{sec:Results}

Adopting the continuity SFH, the Salim attenuation law, and the BC03 stellar templates as our baseline model, we explore the distributions of the derived host galaxy properties (Sec \ref{subsec:host_galaxy_properties}), the inferred DTD from the baseline model (Sec \ref{subsec:baseline_dtd}), how other SED modelling choices can change the inferred DTD including the power law index and how to choose the best model (Sec \ref{subsec:dtd_model_variations} --\ref{subsec:model_selection}). Finally, we study the inferred progenitor ages from the convolution of the DTD with the SFHs of the hosts (Sec \ref{subsec:ages}).

\subsection{Host Galaxy Properties}
\label{subsec:host_galaxy_properties}

\begin{figure}
    \centering
    \sidecaption
    \includegraphics[width=\linewidth]{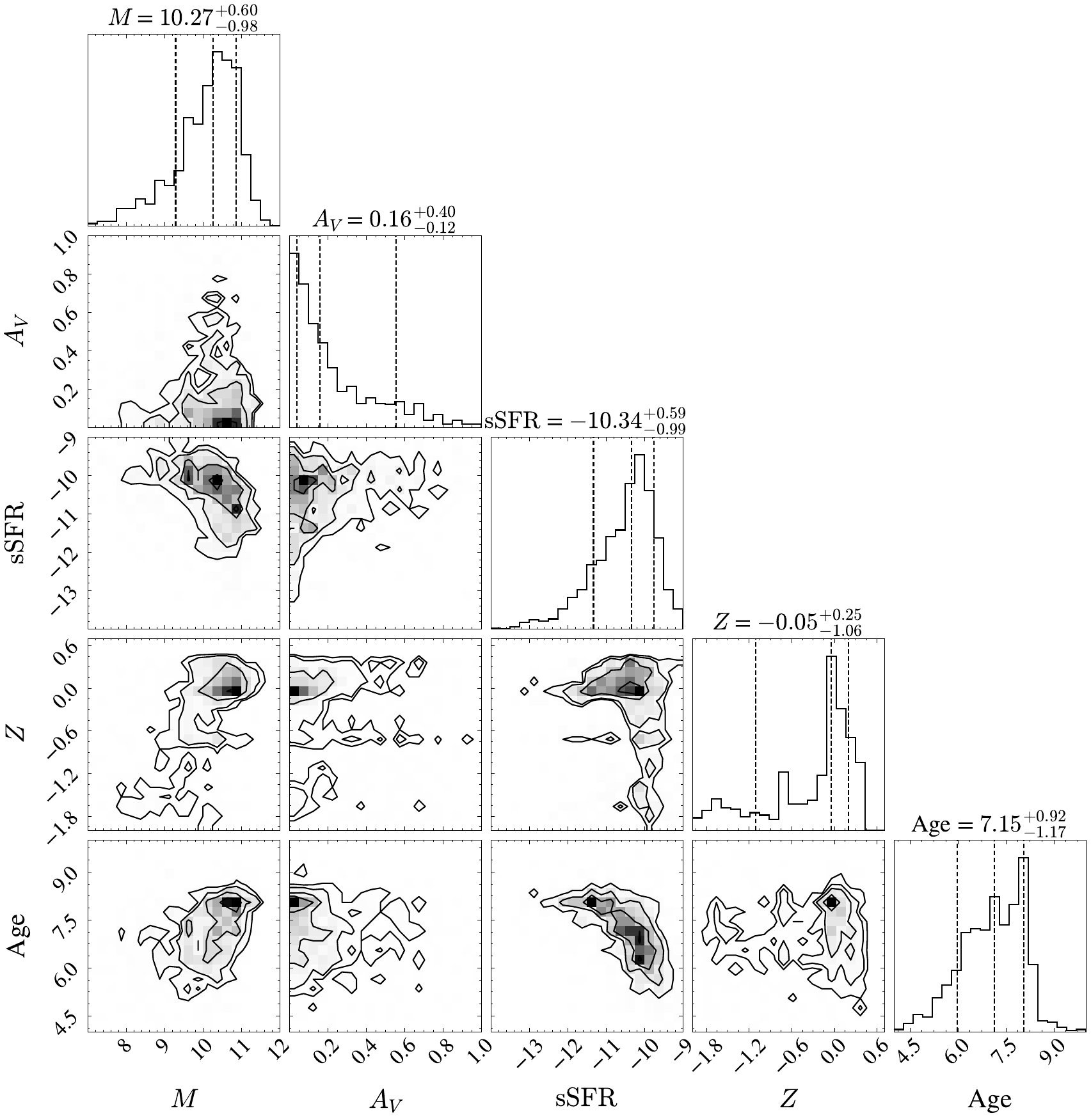}
    \caption{Distributions of the median host galaxy posteriors in our baseline model: (a) log stellar mass in $\mathrm{M}_\odot$, (b) dust attenuation $A_V$, (c) log specific star formation rate in $\mathrm{years}^{-1}$ and (d) the stellar mass-weighted age in Gyr. The dashed lines for each property correspond to the 16-50-84th percentile posterior.}
\label{fig:host_galaxy_distributions}
\end{figure}

Figure \ref{fig:host_galaxy_distributions} shows that the distribution of host galaxy masses in our sample range from about $10^7 \mathrm{M_\odot}$ to $10^{12} \mathrm{M_\odot}$. The distribution peaks around $10^{10.3} \mathrm{M_\odot}$. The majority of the host galaxies have near solar stellar metallicity. There are also a number of host galaxies that have sub-solar or super-solar metallicity. The $A_V$ distribution peaks around 0.1 magnitude, with an extended tail. The distribution of the mass-weighted age of the host galaxies ranges from around 5 to 8 Gyr.

We also quantify the different degrees of star formation in the host galaxies. We adopt the classification of \cite{Nugent_Host_classification} \citep[itself from][]{Tachella_Star_forming}. The measurement of star formation comes from the quantity $\mathcal{D}(z)$ which we determine by a combination of the sSFR and the Hubble time at the redshift of the host galaxy, $\mathcal{D}(z) = \mathrm{sSFR} \times t_{H}(z)$. We classify host galaxies with $\mathcal{D}$ > 1/3 as actively star forming, host galaxies with $\mathcal{D}$ < 1/20 as quiescent, and host galaxies with 1/20 < $\mathcal{D}$ <1/3 as transitioning between star forming and quiescent. We find that about 62\% of the host galaxies are star forming, 24\% are transitioning and 14\% are quiescent (as compared to the \cite{Nugent_Host_classification} fractions of 64\%,8\%, and 28\%). The difference in fractions is likely due to the different composition of our samples.

\subsection{Baseline Model DTD}
\label{subsec:baseline_dtd}

\begin{table}[]
    \centering
    \caption{Integral of the DTD across each delay time bin}
    \begin{tabular}{|c|c|}
        \hline
        \textbf{Delay Time (Myr)} & \textbf{Integrated DTD} \\
        \hline
        30-100 & $0.02 \pm0.01$\\
        \hline
        100-330 & $0.62 \pm 0.06$\\
        \hline
        330-1100 & $0.05 \pm 0.02$\\
        \hline
        1100-3600 & $0.16 \pm 0.06$\\
        \hline
        3600-11700 & $0.13 \pm 0.02$\\
        \hline
        11700-13350 & $0.01 \pm 0.01$\\
        \hline
    \end{tabular}
    \label{tab:dtd_integral}
\end{table}

\begin{figure}
    \centering
    \includegraphics[width=\linewidth]{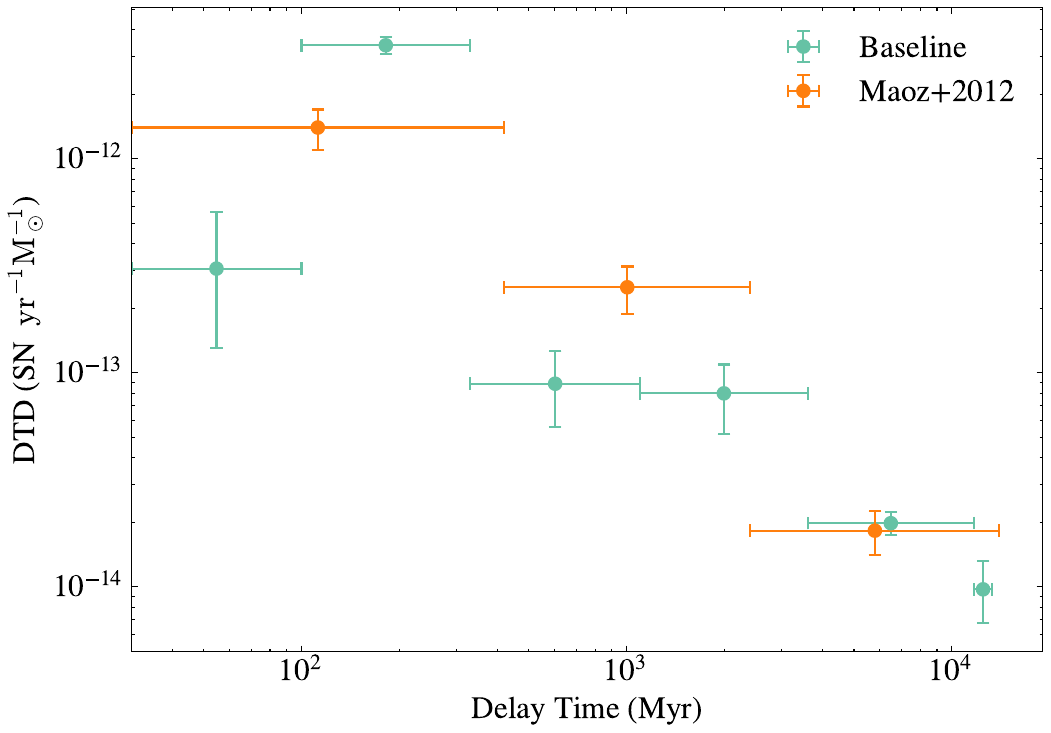}
    \caption{Comparison of the DTD from our baseline model to \cite{Maoz_2012}}
    \label{fig:standard_dtd}
\end{figure}

We normalize our baseline DTD to $\eta = 0.00130\ \mathrm{SN} \ \mathrm{M_\odot}^{-1}$  as in \cite{Maoz_2012} and compare our DTD to the \cite{Maoz_2012} DTD in Figure \ref{fig:standard_dtd}. We also show the full posterior of the baseline DTD in Figure \ref{fig:dtd_cornerplot}. Our baseline DTD exhibits a peak between 100--330 Myr with a plateau between 330--3.6 Gyr. Table \ref{tab:dtd_integral} shows $\phi_j \times dt_j$, which is the integral of the DTD in each delay bin. This is equivalent to the fraction of SNe Ia, produced by a unit mass of star formation, that occurs in each delay bin. About 62\% of SNe Ia occur between 100--330 Myr after the formation of their progenitors. There is another period of SNe Ia formation with 16\% between 1.1--3.6 Gyr, followed by 13\% of SNe Ia between 3.6--11.7 Gyr of delay time from progenitor formation. If the DTD had followed a $t^{-1}$ power law distribution, the fractions across all bins would be roughly equal (due to the delay time bins having a roughly equal separation in logarithmic delay time).

An A+B model \citep{Mannucci_A_B} predicts that SNe Ia rates are described by the current SFR and the total mass of the host galaxy. We find that only a small fraction of SNe Ia occur before 100 Myr (which is the typical timescale for SFR calculations). Thus, our findings agree with those of \cite{Raskin_Prompt_Delayed}, where the instantaneous SFR of the galaxy is correlated with, but not the driver of, the prompt channel.

When compared to \cite{Maoz_2012}, our standard DTD is higher before 330 Myr, lower until 3.6 Gyr, and has similar values at longer delay times. In Paper II, we show that this is largely driven by the effect of host stellar metallicity on the DTD. 

\subsection{Variation of DTD with SED model parameters}
\label{subsec:dtd_model_variations}

\begin{figure*}
\sidecaption
    \centering    \includegraphics[width=10cm]{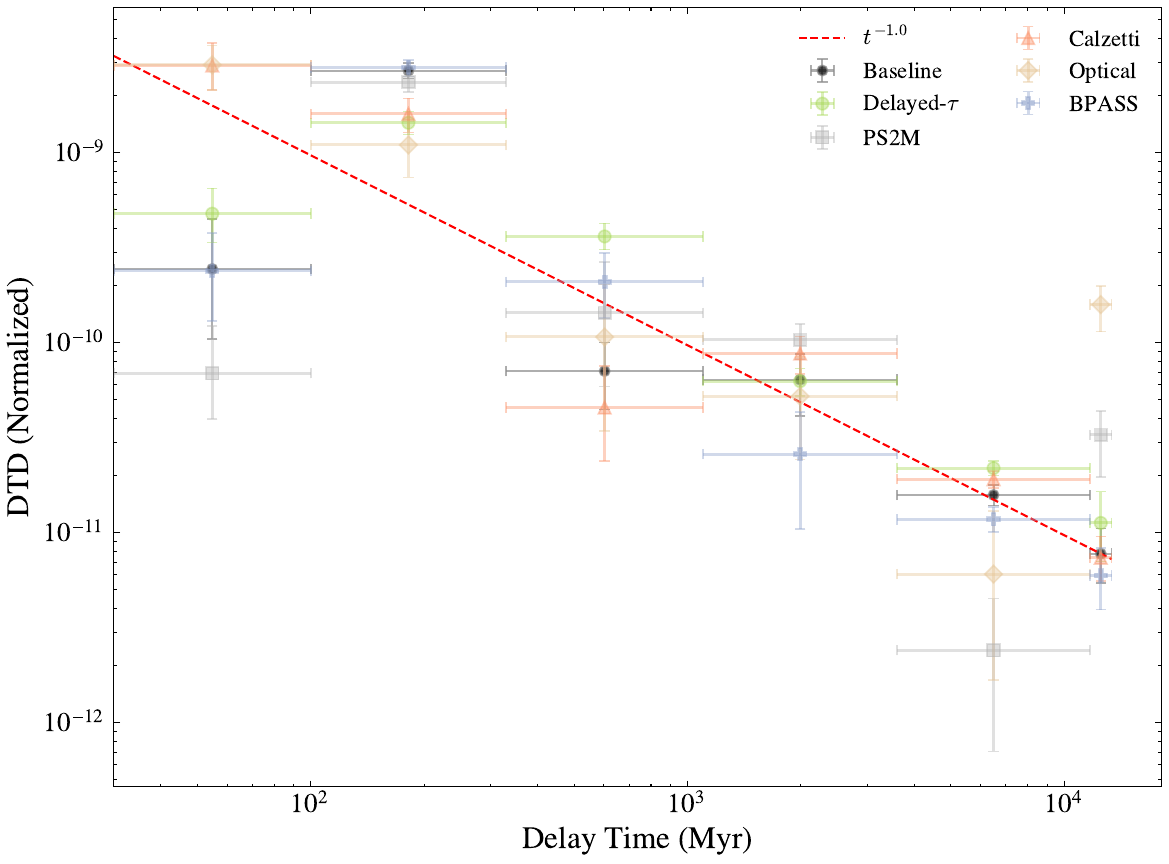}
    \caption{A comparison of the baseline DTD with variations in stellar population templates (BPASS), dust attenuation law (Calzetti), SFH model (delayed-$\tau$), and photometric coverage (optical, PS2M). A $t^{-1}$ power law derived from the baseline DTD value in the final bin is marked with a red dashed line.}
    \label{fig:dtd_comparison}
\end{figure*}

One important source of differences in any framework that infers DTD from the SFH of host galaxies is the weighting of stellar mass formed in each bin. If the SFH assumed during SED modelling creates a similar amount of mass formed in a certain delay time bin across different host galaxies, the DTD will have a higher value in that bin. The continuity SFH can infer more general SFHs than a specific functional form. We can also constrain the SFH better with a wider wavelength range from UV to NIR. We plot the comparison of the DTDs in Figure \ref{fig:dtd_comparison}.

\subsubsection{Effects of the SFH on the DTD}

The delayed-$\tau$ DTD resembles a $t^{-1}$ power law. Appendix \ref{app:parametric} shows that this is driven by the functional form of the delayed-$\tau$ SFH, which does not model quiescent galaxies well. This is consistent with \cite{Parametric_SFH}. The constant star formation rate across recent bins from the delayed-$\tau$ model in quiescent galaxies creates the $t^{-1}$ power law DTD. The continuity SFH prior favours continuous SFRs across different time bins \citep{Leja2019}. The fact that the continuity SFH is able to identify quiescent galaxies that have lower SFR at shorter lookback times indicates that it is not strongly prior-dependent. Thus, we conclude that the delayed-$\tau$ SFH is biased towards the $t^{-1}$ power law as it overestimates the similarity of recent SFRs between quiescent and star forming. \cite{Parametric_SFH} and \cite{Leja2019} also found that the delayed-$\tau$ model along with the other parametric models tend to misrepresent the hierarchical star formation histories of a sample as compared to cosmic star formation history.

\subsubsection{Effects of dust model on the DTD}
\label{subsubsec:dust_dtd}

The DTD from both the Calzetti and the baseline Salim dust extinction laws show no significant differences beyond the first delay time bin (30--100 Myr). This difference likely stems from the fact that the Calzetti extinction law can predict less reddening at a given $A_V$ than the Salim extinction law which has a flexible attenuation slope. This causes a bias to SFHs with lower SFR in the first delay bin to model the observed reddening across our sample. In turn, this makes the recent star formation more similar across galaxies which increases the value of the DTD in the first delay bin. We will study the resulting reddening and attenuation in detail in Paper III.

\subsubsection{Effects of the photometric coverage on the DTD}

Finally, we compare the difference in DTDs derived from data that uses different photometric data for the host galaxies. The DTD derived from purely PanSTARRS data (the optical DTD) exhibits the same behaviour at delay times less than 1.1 Gyr as the delayed-$\tau$ DTD, but with larger error bars. This is likely due to the SFH of the quiescent host galaxies being biased to actively star forming forms with the lack of NIR data providing information about old stellar populations. However, the large error in the optical DTD also indicates that it is not well constrained.

The combination of PanSTARRS data with NIR data from 2MASS makes the PS2M DTD more consistent with the baseline DTD in most delay bins. However, the DTD is not well constrained in the 3.6--11.7 Gyr delay time bin. The baseline model is able to constrain the DTD in this bin with the addition of WISE and GALEX data.

\subsubsection{Stellar population template choice}
\label{subsec:stellar_template}

The difference in the BPASS and BC03 stellar population templates influences each of their inferred DTD. The main difference comes from the behaviour around delay times of 330 Myr--1.1 Gyr and 1.1--3.6 Gyr. The BPASS DTD exhibits a higher value in the 330--1.1 Gyr delay time bin and lower value in the 1.1--3.6 Gyr delay time bin. However, the value at times earlier than 330 Myr and later than 3.6 Gyr is the same as the baseline model. 

To investigate the differences in BPASS and BC03 stellar population templates, we also compare other SED model derived properties in Appendix \ref{app:bpass}. We find that the main difference is that for a given host galaxy, the SED modelling with BPASS stellar templates has solutions that explain the flux in redder photometric bands with higher $A_V$ and lower age and metallicity, while the SED modelling with BC03 stellar templates finds the inverse. We find that the BC03 modelled parameters follow the mass-metallicity relation \citep{Gallazi_mass_metallicity} and observed SNe Ia peak magnitude scatter better than parameters modelled with the BPASS stellar templates.

\subsection{Power law Models}
\label{subsec:power law_models}

We also model our DTD as a power law as is commonly assumed \citep[e.g. ][]{Maoz_2011,Maoz_2012,Wiseman2021,Dwarf_Galaxy_DTD,Titan_SN_DTD}. The start time and characteristic index of an assumed power law DTD has been investigated by these previous studies. From our non-parametric DTD models, we find that the peak of the DTD occurs around delay times of 100--330 Myr, implying a power law start time within that range. Here, we choose to sample $t_\text{min}$ as a free parameter, but find that most models peak in the aforementioned range.

\begin{table}
    \centering
    \begin{tabular}{|c|c|c|}
        \hline
        \textbf{Work/Dataset} &
        $\alpha$ & $t_\text{min}$ (Myr)\\ 
        \hline
        Baseline & $-1.29 \pm 0.04$ & $108^{+13}_{-6}$\\ 
        \hline
        Calzetti & $-1.12 \pm 0.03$ & $66^{+21}_{-15}$\\ 
        \hline
        BPASS & $-1.52 \pm 0.05$ & $116^{+24}_{-11}$\\
        \hline
        Delayed-$\tau$ & $-1.01 \pm 0.03$ & $96^{+3}_{-7}$\\
        \hline
        Optical & $-1.12 \pm 0.06$ & $78^{+15}_{-18}$\\
        \hline
        PS+2M & $-1.55\pm 0.07$ & $248^{+34}_{-56}$\\
        \hline
        \cite{Maoz_2012} & $-1.12\pm 0.08$ & 40 \\ 
        \hline
        \cite{Wiseman2021} & $-1.13\pm0.05$ & 40 \\ 
        \hline
    \end{tabular}
    \caption{Sampled Host DTD power law index $\alpha$ and $t_\text{min}$ recovered under different SED model assumptions.}
    \label{tab:power law_hosts}
\end{table}

Table \ref{tab:power law_hosts} shows the start time is inferred to be in the 100--330 Myr bin as in the non-parametric DTD, with the exception of the DTD inferred by a Calzetti dust attenuation law that peaks at a delay time less than 100 Myr. Our standard model has a power law index of $-1.3$, steeper than most previous studies that find a DTD power law index around $-1.1$. However, when adopting the delayed-$\tau$ SFH, the resulting DTD is consistent with $t^{-1}$ as in previous works \citep{Maoz_2011,Maoz_2012,Wiseman2021}. The optical DTD is also close to the \cite{Maoz_2012} and \cite{Wiseman2021} values of $\sim -1.1$, but this is also likely due to the poor modelling of quiescent galaxies without NIR data. Furthermore, our baseline model is not well modelled with a power law as shown in Figure \ref{fig:standard_dtd}. There are also both theoretical and observational studies that have shown that the DTD might be steeper or composed of a broken power law instead \citep{Ruiter2009,Color_DTD_Heringer,Dwarf_Galaxy_DTD, DTD_Simulation_26}.

The power law index of the DTD with the Calzetti attenuation law is close to the equivalent power law from the non parametric DTD inferred in \cite{Maoz_2012} that used the same attenuation law. It is also similar to \cite{Wiseman2021}. However, \cite{Maoz_2011} compared DTDs inferred with a Calzetti extinction law, as well as a dust extinction law that has separate attenuation for younger and older stars, they found that the main difference was that the value of the DTD in their first delay time bin between 0--420 Myr was lower for the two component dust extinction model, supporting our findings from Section \ref{subsubsec:dust_dtd} that different dust models can affect the bins with the shortest delay time. 

\subsection{DTD Model Comparison}
\label{subsec:model_selection}

We compare the Bayesian model evidence Z computed in \texttt{jaxns} and the expected log predictive density (ELPD) computed using Pareto smoothed importance sampling leave-one-out cross validation method (more simply, LOO) in \texttt{arviz} for each of our DTD models. The evidence can be more sensitive to the choice of priors provided to a model, while the ELPD is less sensitive as it is determined by predicting out of sample data (see \cite{eabm_2025} for a more detailed description of model selection).

The ratio of evidences between two generic models, model 1 and model 2, $Z_1/Z_2$ is known as the Bayes factor between the two. \cite{jeffereys_bayesfactor} defined a scale with a Bayes factor greater than 100 being decisive. \cite{kass1995bayes} defined a Bayes factor of greater than 150 as being very strong (the highest category on their scale). Thus, the difference in the natural logarithm of $Z_1$ and $Z_2$, $\Delta\ln Z > \ln 150 \sim5$ indicates very strong evidence in favour of model 1.

The ELPD quantifies how well the model performs on data outside of its training set. The ELPD is determined through the LOO method by iteratively leaving out observations and finding the predictive performance for using the remaining observations. While explicitly refitting the model on each observation is expensive, \texttt{arviz} uses Pareto smoothing to estimate the LOO without requiring the model to refit each observation. We report the difference in the ELPD between DTD models $\Delta \mathrm{ELPD}$ and $\Delta \text{SE}$, the standard error of the difference in ELPD between DTD models. When $\Delta \mathrm{ELPD} > 4$ and $\Delta \mathrm{ELPD} > \Delta \mathrm{SE}$  between a generic model 1 and model 2, model 1 is favoured \citep{sivula_loo,eabm_2025}.

We only consider DTD models that are derived from the same initial photometric data which excludes those that are derived from different photometric coverage. Out of 969 host galaxies, only 963 galaxies have SED models that have converged through differing SED model choices. We refit all DTD models on these 963 galaxies as the evidence and LOO are both sensitive to the number of data points in the sample. 

\begin{table}[]
    \centering
    \caption{Difference in logarithmic evidence ($\Delta \ln Z$), ELPD ($\Delta \text{ELPD}$), and the standard error on the ELPD ($\Delta \text{SE}$) for DTD models trained on the same data.}
    \begin{tabular}{|c|c|c|c|}
        \hline
        \textbf{DTD Model} & $\Delta \ln Z$ & $\Delta \mathrm{ELPD}$ & $\Delta \mathrm{SE}$ \\
        \hline
        Baseline & 0 & 0 & 0\\
        \hline
        Delayed-$\tau$ & 20 $\pm$ 0.2 & 20 & 12\\
        \hline
        BPASS & 22 $\pm$ 0.2 & 23 & 12\\
        \hline
        Calzetti & 46 $\pm$ 0.2 & 47 & 32\\
        \hline
    \end{tabular}
    \label{tab:model_comparison}
\end{table}

Table \ref{tab:model_comparison} shows the evidence and LOO both favour the baseline model over the other models. $\Delta \ln Z \geq 20$ for the baseline DTD model compared to all the other DTD models, indicating a decisive \citep{jeffereys_bayesfactor} or a very strong \citep{kass1995bayes} preference. $\Delta \mathrm{ELPD}$ is similarly $\geq e^{20}$ in favour of the baseline model, as well as greater than $\Delta \mathrm{SE}$ between the baseline model and every other model. Thus, the baseline model is clearly favoured using both the LOO and the Bayes factor metrics. The ELPD metric explains the performance of the DTD model on data outside of the training sample and is not sensitive to the assumed prior. The Bayes factor metric is sensitive to the choice of the prior in the DTD model but better captures the performance of the DTD model on the training sample (i.e. the whole of the host galaxy sample). Thus, we recommend the baseline model prescription for other SNe Ia DTD studies. The other SED model assumptions have similar values in both the LOO and Bayes factor metric. They are, in order, the DTDs derived from SED models with the delayed-$\tau$ SFH, BPASS stellar population templates, and the Calzetti dust extinction law assumptions in the SED modelling.

\subsection{Progenitor Ages}
\label{subsec:ages}

\begin{figure*}[t]
    \centering
    \begin{subfigure}{0.33\textwidth}
        \centering
        \includegraphics[width=\linewidth]{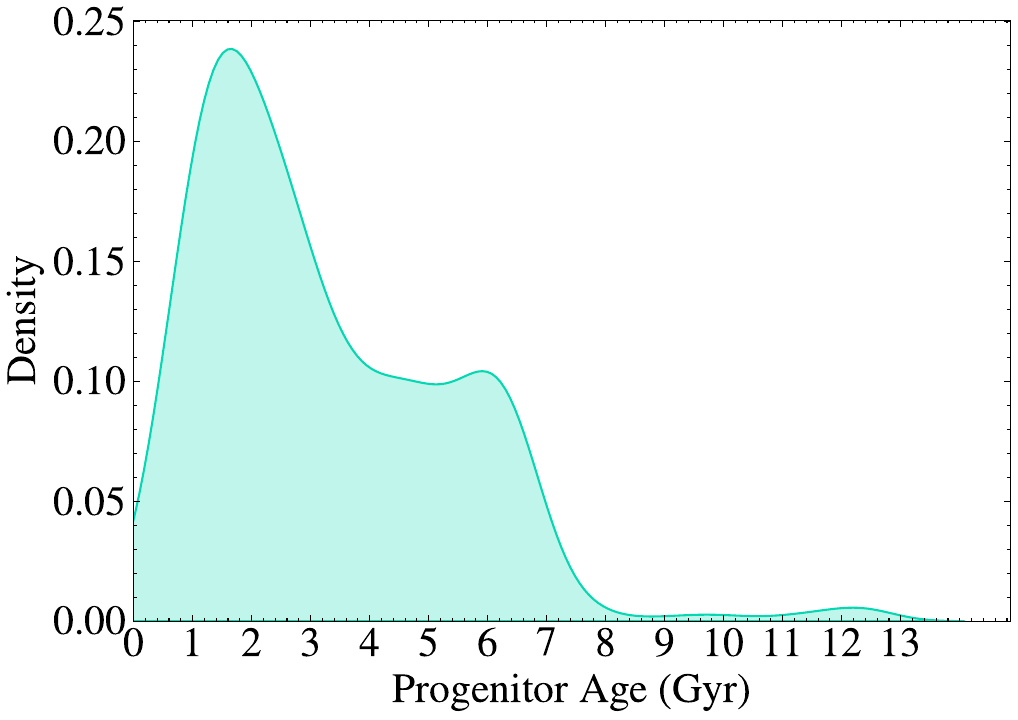}
        \label{fig:age_dist}
    \end{subfigure}
    \hfill
    \begin{subfigure}{0.66\textwidth}
        \centering
        \includegraphics[width=\linewidth]{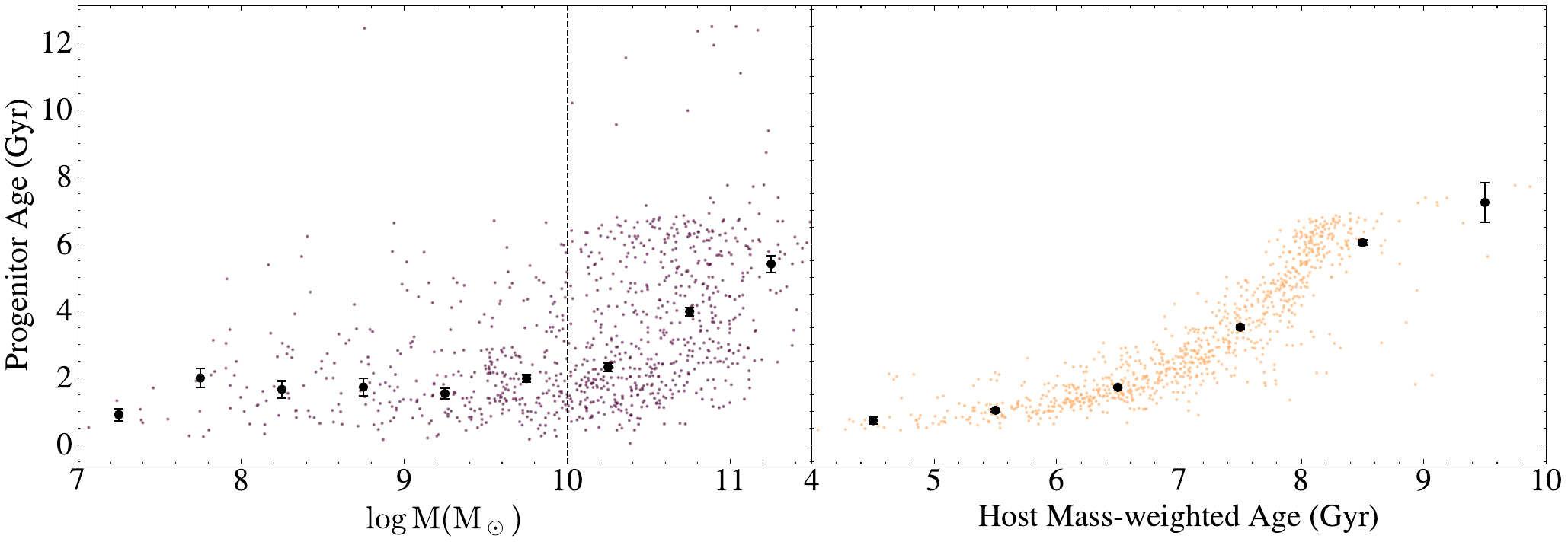}
        \label{fig:mass_age}
    \end{subfigure}

    \caption{Inferred progenitor age distribution (left) and progenitor age against host-galaxy mass (middle) and stellar mass weighted age (right) for our sample. The black dots indicate bins in mass (0.5 dex) and age (1 Gyr). The dashed black line marks $10^{10}\mathrm{M_\odot}$.}
    \label{fig:progenitor_age}
\end{figure*}

Figure \ref{fig:progenitor_age} (left) shows that the distribution of progenitor ages is bimodal, with a prompt peak around 2 Gyr and a delayed peak around 6.5 Gyr. This corresponds to the 1.1--3.6 Gyr bin and the 3.6--11.7 Gyr bins, matching well with the peaks around 2.2 Gyr in star forming galaxies and 6.0 Gyr in quiescent galaxies from \cite{Titan_SN_DTD} inferred from SFH from the TITANS sample modelled with \texttt{dense\_basis} SFH in \texttt{BAGPIPES} and an assumed power law DTD. If we adopt 3.6 Gyr as a cut-off point between the two distributions, we find that approximately 64\% of the progenitors in our sample are younger than the cut-off while 36\% are older. While future papers in the series will examine the relations between age, host, and light curve properties in more detail, we investigate trends in progenitor age with host galaxy mass and stellar mass-weighted age and the SN Ia light curve stretch parameter $x_1$. 

\begin{figure}
\sidecaption
    \centering
    \includegraphics[width = \linewidth]{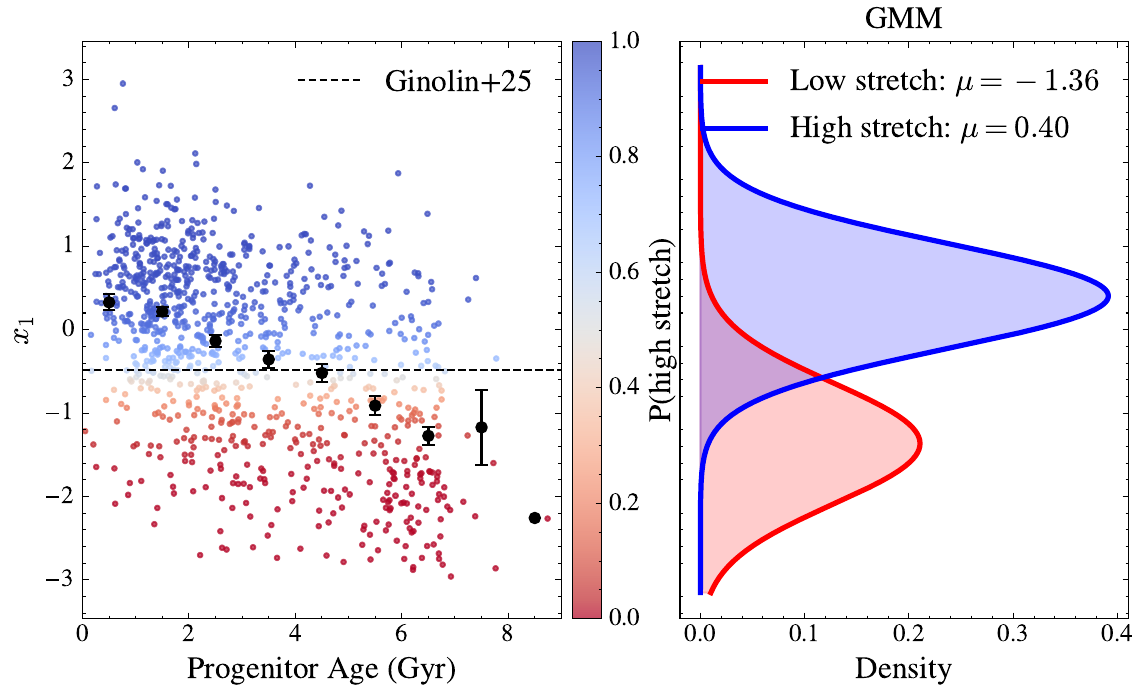}
    \caption{The progenitor age plotted against the light curve stretch ($x_1$) for our sample, with black dots indicating the binned stretch value across 1 Gyr age bins. The right inset shows the two stretch populations and their means. The colour of the background is set by the probability of belonging to the high stretch population}
    \label{fig:stretch_age}
\end{figure}

From Figure \ref{fig:progenitor_age} (middle), a clear evolution of progenitor age occurs around $10^{10} M_\odot$, where hosts transition from purely younger to a mix of young and old progenitor ages. The mass scale of this transition approximately coincides with the commonly used host galaxy stellar mass for the mass step correction \citep{Kelly_2010_Hosts,DLR_Paper_1,Ginolin_Stretch_Env, DES_Mass_Step}. Figure \ref{fig:progenitor_age} (right) shows that the progenitor age and the age of the host galaxy correlate, but are not equal. Younger galaxies tend to host younger SNe Ia, while old galaxies host SNe Ia whose progenitor ages correspond to the older delay time tail of the DTD. This is similar to the findings in \cite{Childress_2014} where younger SNe Ia occur in star forming host galaxies, while the older SNe Ia originate from passive hosts creating a bimodal distribution of progenitor ages with a sharp transition.

Finally, we investigate possible correlations between the age of the SNe Ia progenitor and the SALT 2.4 light curve stretch parameter $x_1$. From Figure \ref{fig:stretch_age}, we find that similar to other studies \citep{Sullivan_2006,Bradt2010,Pan_PTF_paper,Rigault_lsfr}, high stretch SNe Ia are mainly associated with younger progenitors while low stretch SNe Ia are mainly associated with older progenitors . In \cite{Ginolin_Stretch_Env}, the cut between high and low stretch populations is found to be roughly $x_1 = -0.48$, which we indicate with a dashed line.

We fit a Gaussian mixture model (GMM) to the stretch distribution similar to \cite{Ginolin_Stretch_Env} and \cite{Radek_ZTF_two_pop} and find that they make up two populations, one high $x_1$ and one low $x_1$. The GMM weights of each population are around 62\% in the high stretch population and 38\% in the low stretch population, aligning well with the fractions of younger and older progenitors with a cut at 3.6 Gyr. This is also in line with \cite{Ginolin_Stretch_Env} with $59\pm7$\% in the high stretch mode. We find a slight difference with \cite{Radek_ZTF_two_pop} with $52\pm4\%$ but this is due to \cite{Radek_ZTF_two_pop} also accounting for differences in intrinsic brightness, colour, and host properties.

\section{Discussion}
\label{sec:Discussion}

We infer the presence of two progenitor age populations that correlate with two population in SNe Ia light curve stretch $x_1$. Both \cite{Ginolin_Stretch_Env} and \cite{Radek_ZTF_two_pop} have found that the high stretch population should be standardized separately from the low stretch population, particularly with the stretch coefficient $\alpha$. \cite{Radek_ZTF_two_pop} and \cite{Rubin_2026_lum_gap} both find a luminosity gap between these two populations. \cite{Rojas_Bravo_Bump} find that early-phase flux excess in the ZTF sample also shows a correlation with higher $x_1$.  We have shown that there is a difference in progenitor age between these two population that could be the underlying cause of the different behaviour of these populations. However, we also caveat this by noting that while the high stretch SNe Ia dominate younger ages, there are still a number of low $x_1$ SNe Ia (while the high $x_1$ SNe Ia are almost absent at progenitor ages older than 3.6 Gyr).

However, one of the fundamental assumptions that we have adopted in this work is that the DTD is a universal function for all the hosts in both shape and amplitude. All of our DTD models show higher SNe Ia rates at delay times less than 330 Myr. However, our inferred age distribution finds ages much larger than this prediction. Only 9 SNe Ia from our sample come from progenitors younger than 330 Myr. This could indicate that the observed SNe Ia in our sample all originate from progenitors older than 1--2 Gyr. However, this is unlikely as most host galaxies in our sample are actively star forming.

Alternatively, it is likely that the DTD itself is not a universal function. A DTD composed of multiple progenitor channels, with one channel having high SNe Ia rates at short delay times and minimal SNe Ia at longer delay times, and another channel that exhibits significant SNe Ia rates with a more extended range of delay times could still be inferred by the universal method, but would fail to recover accurate ages. SNe Ia generated by the prompt channel in star forming galaxies will have their inferred age biased by the stellar mass in these galaxies that formed at long lookback times. This would bias inferred age estimates higher for these SNe Ia, potentially explaining the difference in progenitor age and DTD peak in our baseline model. This was noted in \cite{Nonparametric_DTD}, who found that the age of the host could be a systematic in the inferred DTD. Different channels can overcome some of these systematics caused by tracing star formation and host ages. The alternative approach would be to find the basis vectors of the channels embedded in the universal DTD and how much they each contribute. This is the subject of the following paper in this series.

The other major assumption is the normalization of the DTD. A universal DTD assumes an absolute normalization. However, recent studies have found that the normalization might be dependent on the metallicity of the formation environment \citep{Gandhi_DTD_Metallicity,Johnson_Metallicity_DTD,Burgaz_low_mass_hosts,Dwarf_Galaxy_DTD, DTD_Simulation_26}. Our sample spans a wide range of host galaxy stellar metallicities from sub-solar to super-solar, possibly biasing the recovered DTD. The dependence of the normalization on the host stellar metallicity will also be the subject of study in future papers. These two factors taken together indicate that our results should be interpreted with caution. We have recovered the DTD only up to the assumption that the DTD is a universal function.

Our study also has implications for future DTD modelling from large surveys such as the LSST \citep{LSST_Overview}. The DTD recovered from hosts with purely PanSTARRS photometry (which matches LSST wavelength coverage) is not sufficient to constrain the DTD. It is recommended that additional infrared photometry from facilities, such as the Nancy Grace Roman Space Telescope \citep{Roman_Paper} and Euclid \citep{Euclid_Paper}, is utilized to better model the SED, and hence the DTD. Data from mid to far-infrared is needed for a complete study of host galaxies. This is more significant in the case of higher redshift SNe Ia hosts, where the rest frame wavelength range captured by these surveys will be narrower. 

While local stellar environments can constrain the star formation history of the progenitor better than the global host environment, this also implies that results from local SED modelling could bias the inferences if there is no resolved local NIR photometry. We find that 2MASS and WISE local photometry ($\sim1$--$2$ kpc regions) does not provide significant photometric information due to the low resolution of the surveys even in the low redshift ZTF sample. For this reason, we do not perform a similar analysis on the local environment even though \texttt{HostPhot} is capable of measuring photometry.

We encourage the application of \texttt{CAALAM} on other SN Ia samples to compare inferred DTD, particularly those at lower redshifts with more host photometric coverage. The TITANS DR1 sample \citep{Titans_DR1_sample} has derived host properties in a similar manner as our baseline model, making it a natural sample for \texttt{CAALAM}. We already find similar progenitor ages to \cite{Titan_SN_DTD} with the ZTF sample. Future papers in this series will show how progenitor channels can be inferred from \texttt{CAALAM} and analysed alongside light curve properties.

Finally, our method implemented in \texttt{CAALAM} is not necessarily restricted to SNe Ia. The non-parametric DTD in this study has a prior that favours a power law model, but simply modelling the DTD as a Gaussian can recover DTDs that are more realistic for core-collapse SNe. In fact, any sample of extragalactic astrophysical transients with a well defined selection function and properly modelled host galaxy SFHs can have a hierarchical DTD inferred through \texttt{CAALAM} without the need for SFHs from galaxies that did not host these transients. The inferred ages for these events can also aid in classification in cases with poorly sampled light curves. 

\section{Conclusion}
\label{sec:Conclusions}
We have developed \texttt{CAALAM}, a conditional extension of the \cite{Maoz_2011} method of modelling the DTD of a sample of transient host galaxies from individual SFHs. This method recovers a probability density function of delay times from a given sample of SFHs binned into discrete lookback time bins. Our key findings are summarized below.

\begin{enumerate}
    \item We find that our baseline model with BC03 stellar templates, a \cite{Salim_Dust} dust model, a \cite{Leja2019} continuity SFH model, and photometric coverage, is favoured over models with variations in any of these aspects.
    \item Our inferred DTD implies that around 60\% of all SNe Ia occur within 330 Myr of progenitor formation.
    \item When we model our baseline DTD with a power law, we find that the index is $-1.29 \pm0.04$, which is steeper than previous predictions. When we use a \cite{Calzetti_dust} dust attenuation model or only Pan-STARRs photometry, we find an increase in the DTD between 30--100 Myr, corresponding to a power law index of $-1.12\pm0.03$, in line with the inferred power law indexes of \cite{Maoz_2012} and \cite{Wiseman2021}.
    \item Our baseline model predicts a bimodal distribution of observed progenitor ages with close to two-thirds of SNe Ia occurring within 1.1--3.6 Gyr of progenitor formation. The remaining third mostly originate from progenitors between 3.6--11.7 Gyr old. The younger progenitor SNe Ia produce light curves with higher stretches, while those with older progenitors produce almost exclusively low stretch light curves.
\end{enumerate}

We will release the code associated with \texttt{CAALAM} as well as the baseline model host SED parameter catalogue soon after the publication of all the papers in this series.

\begin{acknowledgements}
      This work was supported by a research grant (VIL54489) from VILLUM FONDEN and a Villum Experiment grant (VIL69896). A.N.R. would like to thank Danial Langeroodi, William Baker, Aidan Sedgewick, and Mart\'in Solar for insightful discussions, as well as the authors of the open source software and all astronomical data used in this study. A.N.R. also thanks Francesca Borg for helpful comments that improved this paper. A.N.R. acknowledges that generative AI (ChatGPT and Claude) was used as an aid to develop the underlying code and as a tool to discuss results but not to generate text used in this paper.
\end{acknowledgements}

%

\bibliographystyle{aa}
\bibliography{references}







   
  



\begin{appendix}





\section{Testing the conditional DTD model} \label{app:conditional_dtd_test}

In this appendix, we explore the behaviour of the host-only DTD model when no signal is present in the data. We accomplish this by randomly shuffling the SFH posteriors across galaxies and samples. 

\begin{figure}
    \centering
    \includegraphics[width=\linewidth]{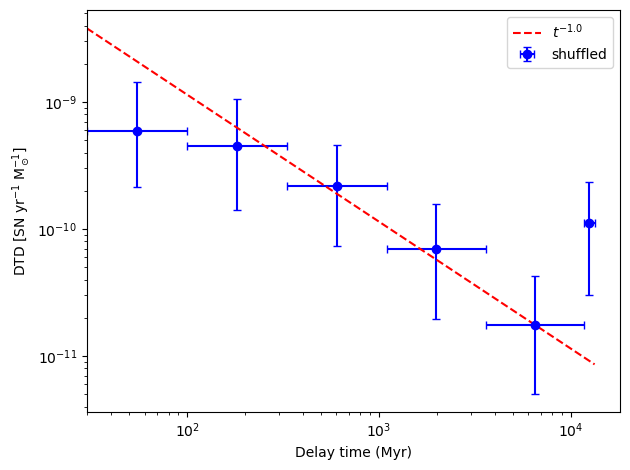}
    \caption{modelled DTD from shuffled SFHs.}
    \label{fig:dtd_shuffle}
\end{figure}

\begin{figure}
    \centering
    \includegraphics[width=1\linewidth]{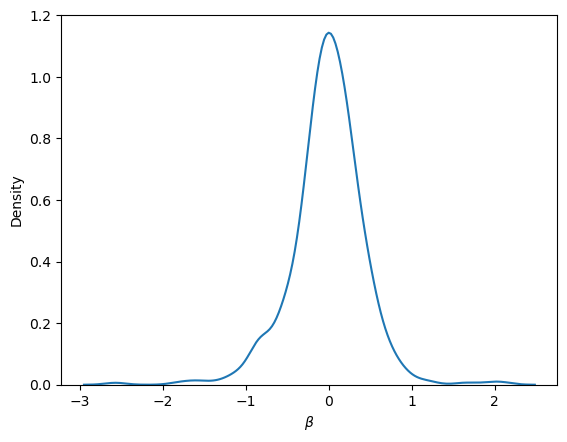}
    \caption{Sampled power law index $\beta$ for the shuffled SFHs}
    \label{fig:dtd_shuffle_power}
\end{figure}

Figure \ref{fig:dtd_shuffle} shows that the DTD inferred from shuffled data without realistic SFHs is driven by the prior. This is set by the Dirichlet prior on equal probability in each bin (as opposed to equivalent rate in each bin), equivalent to a $t^{-1}$ power law as it implies an equal probability in each logarithmic bin. The error-bars associated with each bin are also large and overlap with adjacent bins. The offset of the 30--100 Myr and last bin from the power law likely is due to their binning that deviates from the logarithm. Figure \ref{fig:dtd_shuffle_power} shows the equivalent power law model is strongly peaked around 0, finding that a flat DTD describes the shuffled data best. 

\begin{table}
    \centering
    \begin{tabular}{|c|c|c|c|c|c|c|c|}
        \hline
        \textbf{Archetype} & \textbf{\%} &
        \textbf{$\alpha_1$} &
        \textbf{$\alpha_2$} &
        \textbf{$\alpha_3$} &
        \textbf{$\alpha_4$} &
        \textbf{$\alpha_5$} &
        \textbf{$\alpha_6$} \\ 
        \hline
        Young & 10 & 0.15 & 0.50 & 0.25 & 0.07 & 0.02 & 0.01 \\
        \hline
        Intermediate & 10 & 0.05 & 0.15 & 0.35 & 0.30 & 0.12 & 0.03 \\
        \hline
        Old & 25 & 0.01 & 0.02 & 0.05 & 0.15 & 0.50 & 0.27 \\
        \hline
        Burst & 30 & 0.02 & 0.08 & 0.55 & 0.25 & 0.08 & 0.02 \\
        \hline
        Quenched & 5 & 0.01 & 0.01 & 0.03 & 0.10 & 0.40 & 0.45 \\
        \hline
        Green Valley & 20 & 0.02 & 0.08 & 0.20 & 0.35 & 0.30 & 0.05 \\
        \hline
    \end{tabular}
    \caption{Dirichlet SFH archetypes and their corresponding mass fractions in the six lookback-time bins.}
    \label{tab:dirichlet_basis_random}
\end{table}

We also investigate the effect of randomly sampling a large number of simulated galaxy SFHs. We simulate one million SFHs following six archetypes that are young and star forming, intermediate and main sequence, old and massive, post starburst, quenched, and green valley transitioning. Table \ref{tab:dirichlet_basis_random} shows the Dirichlet basis for each archetype and the assumed fraction for each. We use a vector sampled from the Dirichlet basis and assign a total mass uniformly logarithmically sampled between $10^{8} - 10^{11.5} M_\odot$. Finally, we draw 1000 random galaxies from this distribution and run the model on their SFHs.

\begin{figure}
    \centering
    \includegraphics[width=1\linewidth]{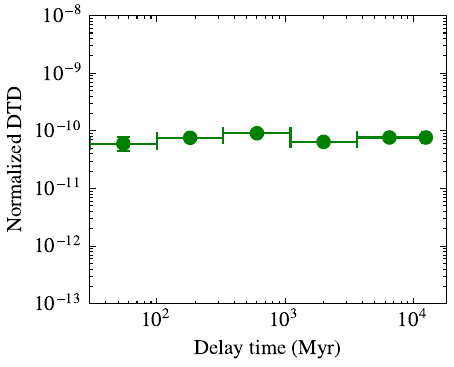}
    \caption{modelled DTD from a large sample of hosts.}
    \label{fig:dtd_large_sample_np}
\end{figure}

Figure \ref{fig:dtd_large_sample_np} shows the recovered DTD from the sample. There are traces of structure present in the DTD. A clear peak in the 330--1100 Myr bin is driven by the unequal fraction of older galaxies in the bin. However, the effect of this structure is minimal. The values are very close to each other and reflect a flat DTD in rate space (they all lie around $7\times10^{-11}$). Thus, with more realistic SFHs, the model finds a flat distribution best describes the data. We also find a similar result with the power law model. 

Thus, we see that a random signal defaults to the prior in the non-parametric model. However, a more realistic simulation defaults to a flat model. The power law model finds a flat model as well in the absence of a signal. As we do not see this behaviour in our fiducial conditional sample, we can conclude that the convergence of the model on real data reflects an underlying signal instead of model bias.

\section{Baseline DTD Posterior}
\label{app:dtd_posterior}

\begin{figure}
    \centering
    \includegraphics[width=\linewidth]{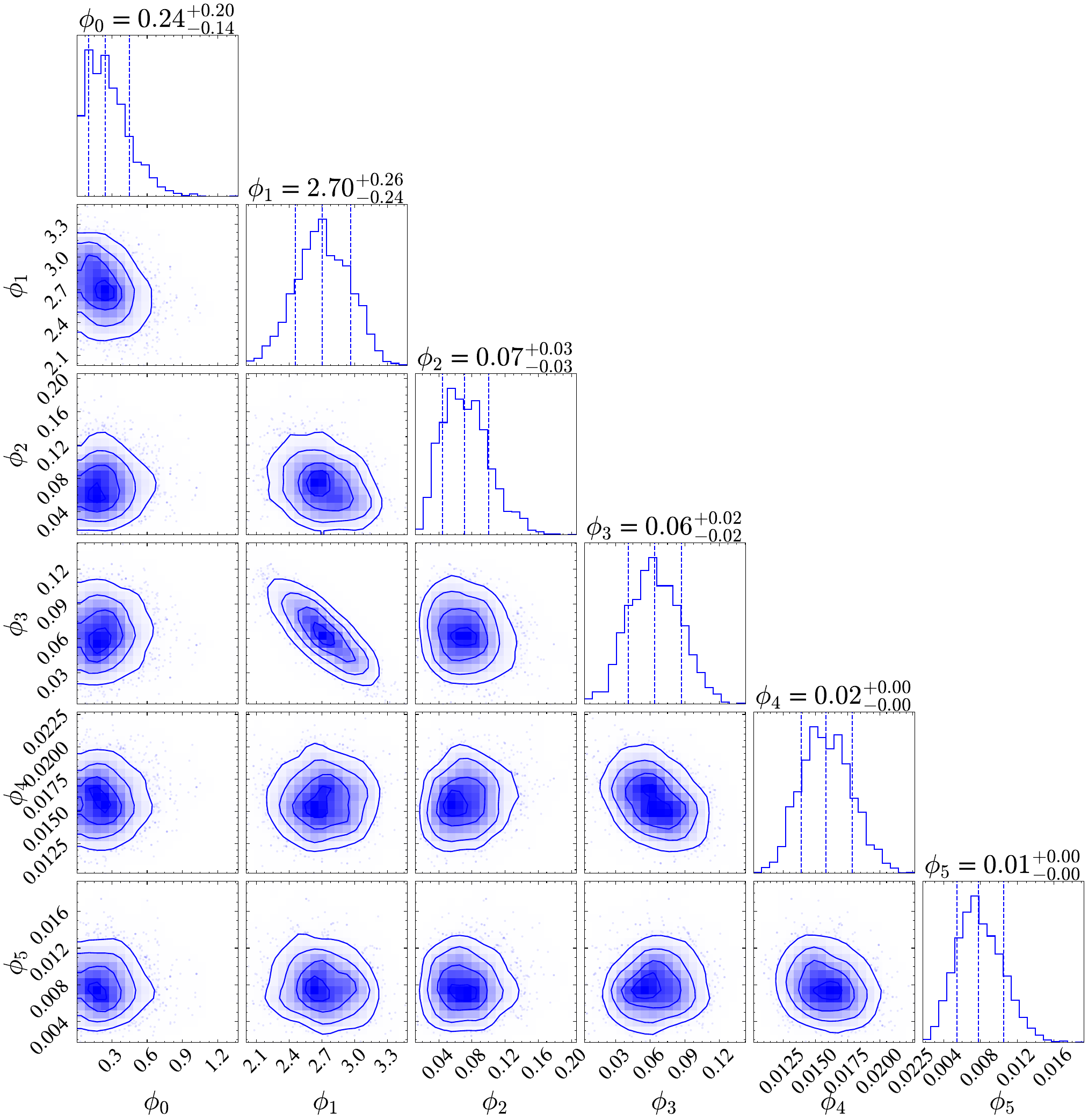}
    \caption{Corner plot of the baseline DTD posterior. The DTD is normalized to 1 and values are multiplied by $10^9$ for better visualization.}
    \label{fig:dtd_cornerplot}
\end{figure}

In Fig. \ref{fig:dtd_cornerplot}, we present the corner plot of the posterior of the baseline DTD. The correlation between the 100--330 Myr bin and 1.1--3.6 Myr bin comes from the geometry of the Dirichlet distribution.

\section{Difference between parametric delayed-$\tau$ SFH and continuity SFH} \label{app:parametric}

\begin{figure}
    \centering
    \includegraphics[width=1\linewidth]{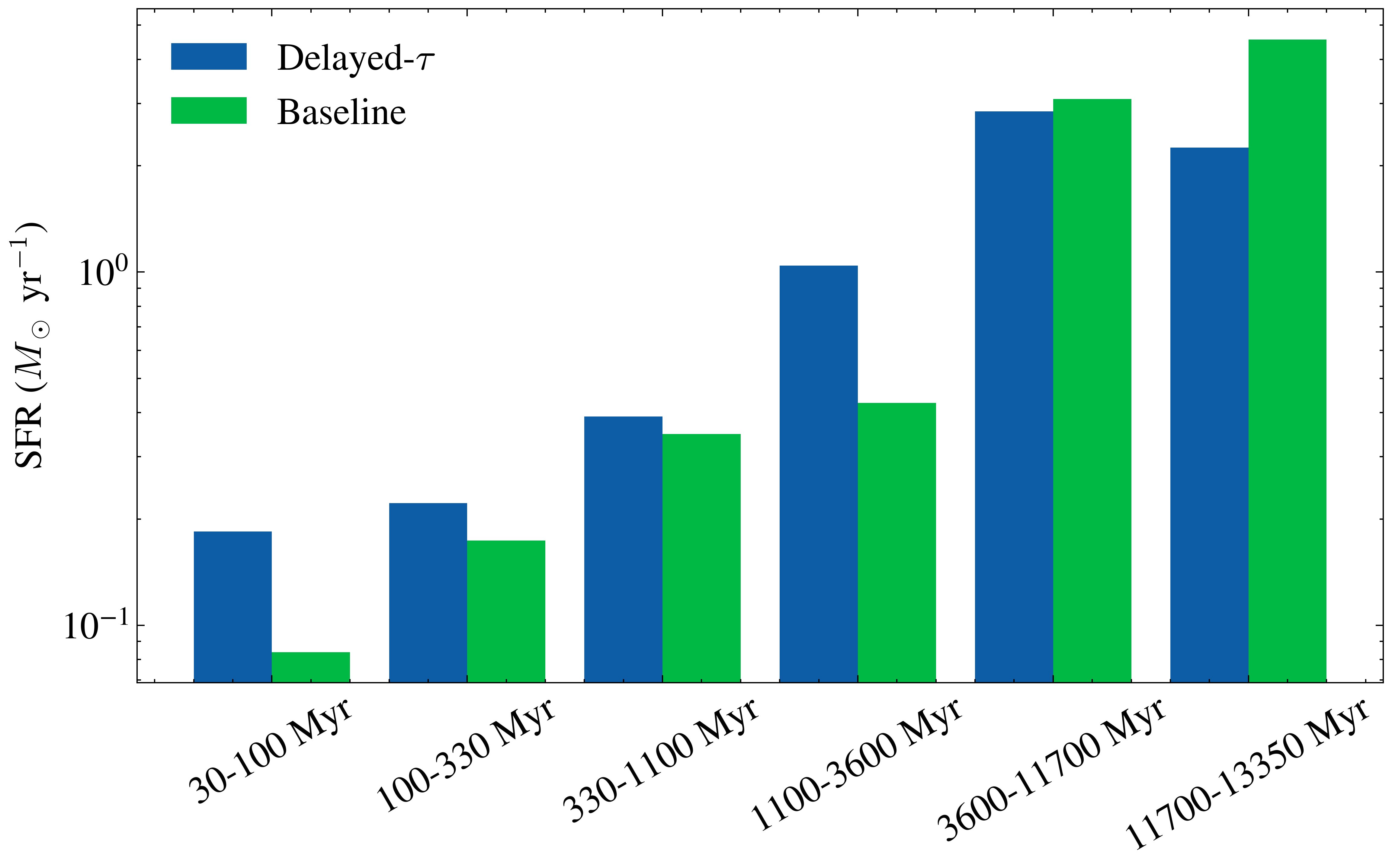}
    \caption{Average star formation rate in each lookback bin of quiescent galaxies (log sSFR < -11) from both the continuity and the delayed-$\tau$ SFH models. }
    \label{fig:parametric_continuity_SFR}
\end{figure}

In order to study the differences between the continuity SFH and delayed-$\tau$ parametric SFH, we study the difference between galaxies labelled as quiescent in our baseline model. Figure \ref{fig:parametric_continuity_SFR} shows the average SFR formed in different lookback bins. We see that the SFR in the most recent bins in the delayed-$\tau$ model is almost constant. The continuity prior on the \cite{Leja2019} SFH favours solutions that tend to have consistent star formation rates in each of the lookback bins. The delayed-$\tau$ model finding more continuous star formation rates across bins as compared to the continuity model indicates that the exponential nature of the model causes star formation to continue in more recent bins even when the data does not favour it in the case of quiescent galaxies. Thus, the continuity prior model in this case does a poor job at modelling quiescent galaxies and creates a bias in the DTD.

\begin{figure}
    \centering
    \includegraphics[width=1\linewidth]{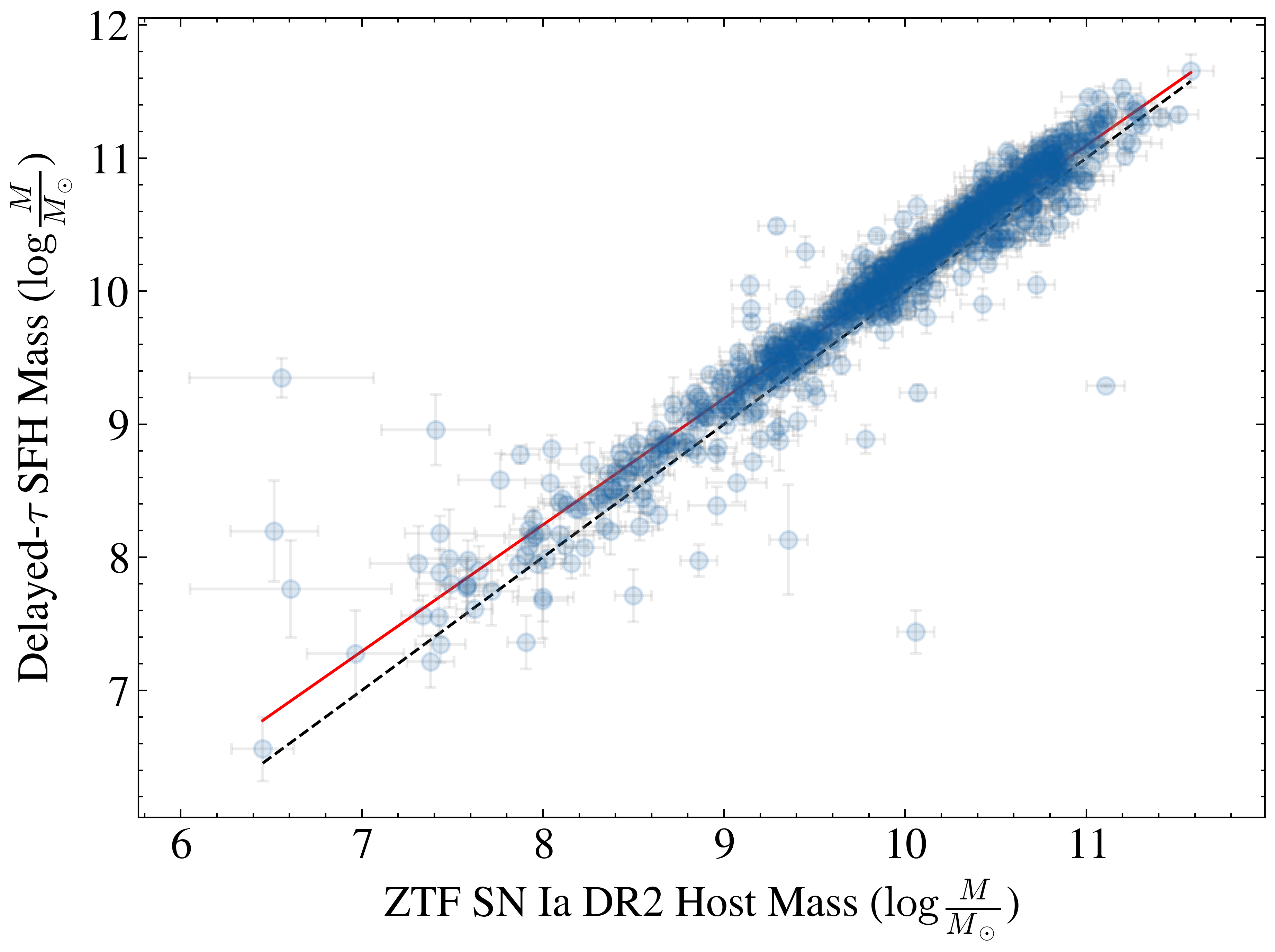}
    \caption{Comparison of stellar masses in the ZTF sample as compared to our work with the delayed-$\tau$ model.}
    \label{fig:ztf_compare}
\end{figure}

We also find that the choice of SFH is the greatest difference between the host stellar masses recovered in our baseline model and the corresponding masses released as part of the ZTF sample. In Figure \ref{fig:ztf_compare}, we see that the delayed-$\tau$ model finds similar masses to the ZTF sample masses with the very low mass galaxies being the main difference. The corresponding linear equation to the best fit line is $y =1.002x + 0.06$ where $y$ our delayed-$\tau$ stellar mass and $x$ is the ZTF sample mass, indicating an offset of only around 0.06 dex between our masses and the ZTF sample masses. This residual is likely driven by the use of supplemental photometric data to PanSTARRS, as well as slight differences in the templates used in our sample.

However, the corresponding equation for the baseline model is $y = 0.969x + 0.482$, indicating a 0.1--0.2 dex offset for galaxies between 9--11 log $M_\odot$. Thus, the differences between our stellar masses and the ZTF sample masses are driven by the choice of the delayed-$\tau$ SFH in the ZTF sample model and the continuity SFH in our baseline model.

\section{Differences between BC03 and BPASS templates} \label{app:bpass}

\begin{figure}
    \centering
    \includegraphics[width=1\linewidth]{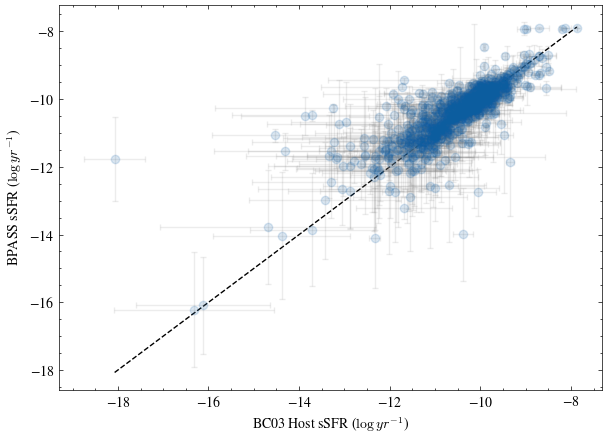}
    \caption{Comparison of modelled SFR between BPASS and BC03. $y=x$ is plotted as a dashed black line}
    \label{fig:SFR_BC03}
\end{figure}

\begin{figure}
    \centering
    \includegraphics[width=1\linewidth]{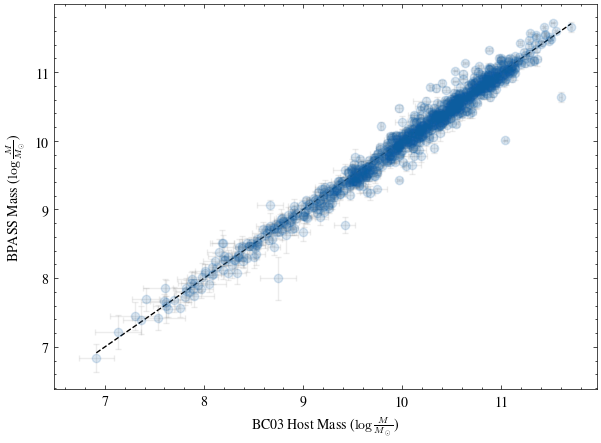}
    \caption{Same as Fig \ref{fig:SFR_BC03} but for stellar masses}
    \label{fig:Mass_BC03}
\end{figure}

\begin{figure}
    \centering
    \includegraphics[width=\linewidth]{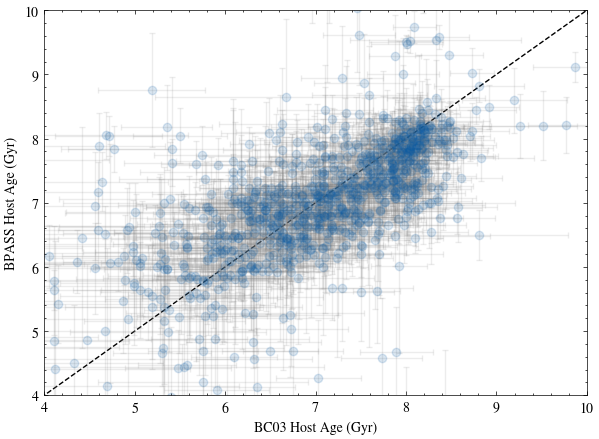}
    \caption{Same as Fig \ref{fig:SFR_BC03} but for the stellar mass-weighted age.}
    \label{fig:age_comparison}
\end{figure}

\begin{figure}
    \centering
    \includegraphics[width=\linewidth]{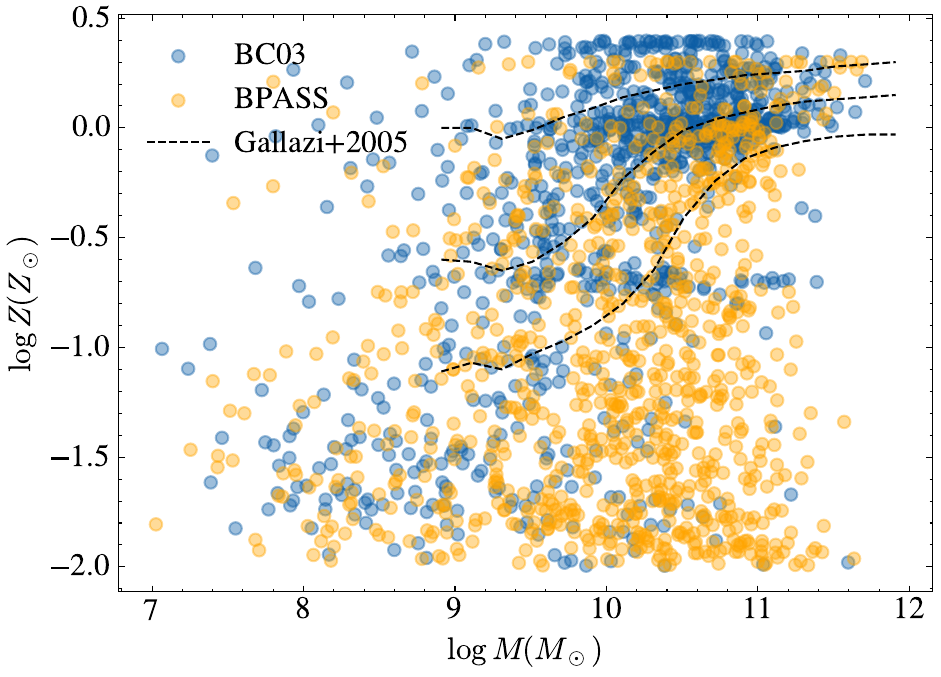} 
    \caption{Comparison between the stellar mass and metallicity of the host galaxies with BC03 (Blue) and BPASS (Orange) templates. The corresponding relation from \cite{Gallazi_mass_metallicity} is plotted in dashed black lines.}
    \label{fig:metallicity_comp_bc03_bpass}
\end{figure}

\begin{figure}
    \centering
    \includegraphics[width=\linewidth]{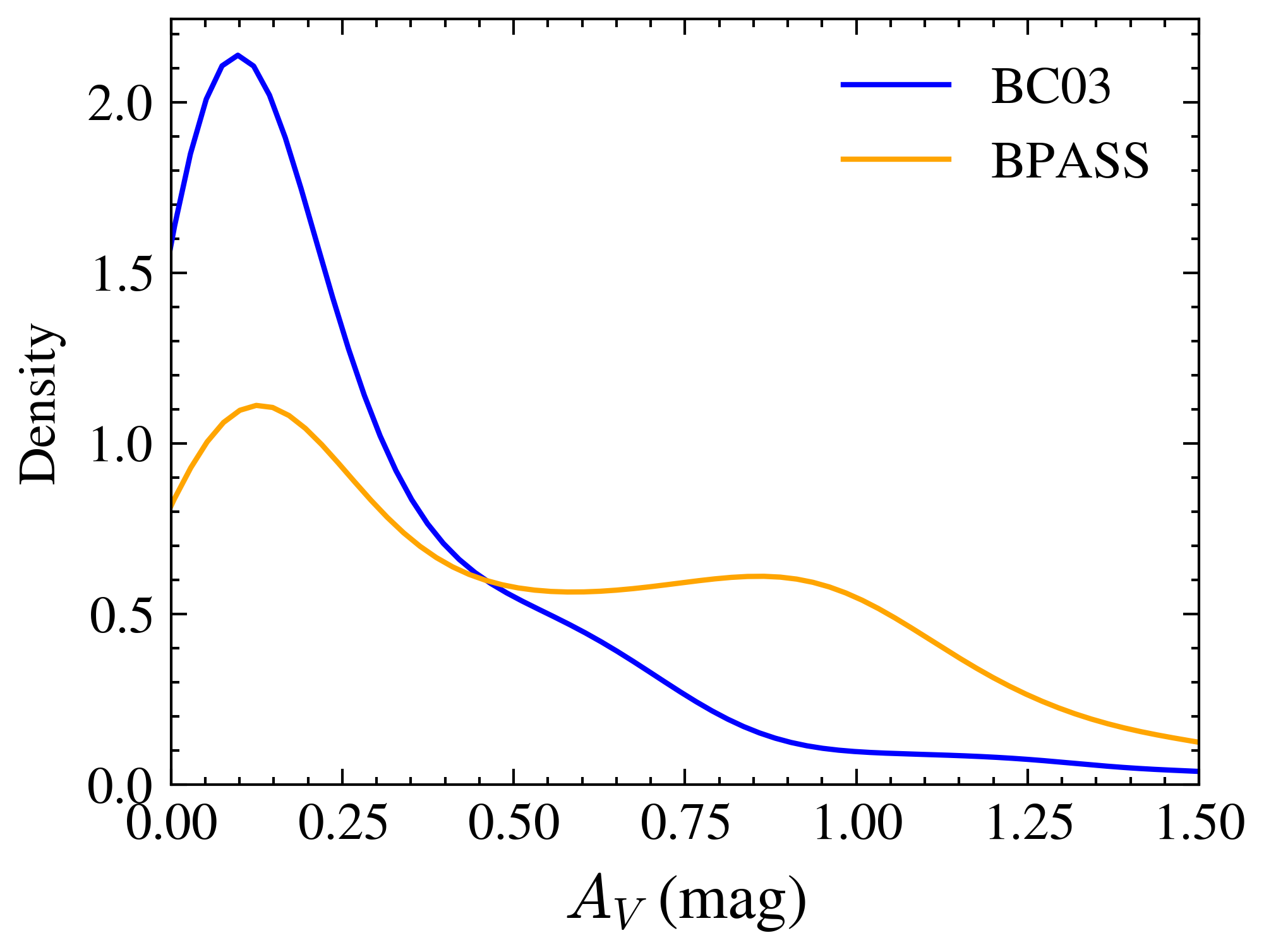}
    \caption{Comparison of the host $A_V$ distributions from BC03 and BPASS stellar templates.}
    \label{fig:Av_comparison}
\end{figure}

In this appendix, we study differences between using BC03 and BPASS models when using Salim dust, the entire photometry and the continuity SFH. (See \cite{Devils_Bpass_BC03} Appendix B and \cite{Bozhidara_galaxy_templates} for other such comparisons)

Figures \ref{fig:SFR_BC03} and \ref{fig:Mass_BC03} show comparisons between the derived stellar masses and sSFR from BC03 and BPASS. The masses agree with each other (similar to \cite{Devils_Bpass_BC03}) with a $y=x$ line being the best fit. However, there are a few systems that have higher sSFR in BPASS as compared to BC03. These are systems that are inferred more passive and likely older in BC03 than in BPASS. This is unlike the trend in \cite{Devils_Bpass_BC03}, where the SFRs aligned at an offset. However, this trend is observed in \cite{Bozhidara_galaxy_templates} at higher metallicities, where BPASS can infer similar or higher star formation rates to BC03.

Figure \ref{fig:age_comparison} shows that the stellar mass-weighted ages derived from BPASS stellar templates are older for galaxies that are younger when the age is derived from BC03 stellar templates, and vice versa. However, the oldest galaxies have the same age using both stellar templates. This aligns with the difference between the 330 Myr--3.6 Gyr delay time bins that is found in Figure \ref{fig:dtd_comparison}. This is a known result from the BPASS vs BC03 templates, where BPASS can predict younger ages due to binary evolution (in line with \cite{Devils_Bpass_BC03}).

Figure \ref{fig:metallicity_comp_bc03_bpass} shows that the metallicity of the BPASS model does not obey the stellar mass-metallicity relation, where the stellar mass and metallicity of the galaxy are correlated \citep{Gallazi_mass_metallicity}. Instead, we find that many high mass galaxies exhibit very low metallicity (around 0.01 $Z_\odot$). While \cite{Devils_Bpass_BC03} found that BPASS templates model lower metallicity than BC03 templates, they should still obey the mass metallicity relation from their work (albeit for gas phase metallicity). The BC03 model does follow the relation for most host galaxies.

Finally, Figure \ref{fig:Av_comparison} also shows that the inferred dust attenuation is larger than that from the BPASS templates. A significant number of host galaxies in the BPASS model have $A_V > 0.5$ mag. This would imply a larger scatter in SNe Ia peak magnitudes than the 0.46 magnitude offset before standardization found among the ZTF sample \citep{ZTF_SN_overview}.

Thus, we find that the main difference is that the BC03 and BPASS stellar templates are accounting for redder SEDs in different ways. BPASS itself contains more bluer flux from the binary populations and thus explains redder SEDs of massive galaxies as younger galaxies with higher dust attenuation and lower metallicity. Meanwhile, BC03 explains the redness with older ages, less dust attenuation and higher metallicity. These changes mainly impact the SFH at the 330--3.6 Gyr timescale. We prefer the baseline model as the metallicities align with standard mass-metallicity relations, the dust aligns with the observed scatter in SNe Ia peak magnitude, and the evidence and LOO metrics favour the baseline model over the BPASS model.

\end{appendix}
\end{document}